\documentclass[referee]{aa}
\usepackage{graphicx}
\begin{document}
\title{Radio, Millimeter and Optical Monitoring of GRB030329 Afterglow: Constraining the Double Jet Model} 
\author{L.~Resmi$^{1,2}$, C.H.~Ishwara-Chandra$^3$, A.J.~Castro-Tirado$^4$, D.~Bhattacharya$^1$, A.P.~Rao$^3$, M.~Bremer$^5$, S.B.~Pandey$^6$, D.K.~Sahu$^{7,8}$, B.C.~Bhatt$^7$, R.~Sagar$^6$, 
G.C.~Anupama$^7$, A.~Subramaniam$^7$, A.~Lundgren$^{9,10}$, J.~Gorosabel$^{4}$, S.~Guziy$^{4,11}$,\\
A.~de Ugarte Postigo$^{4}$, J.M.~Castro Cer\'on$^{12}$ and T.~Wiklind$^{12}$}
\institute{$^1$Raman Research Institute, Bangalore 560080, India\\
$^2$Joint Astronomy Programme, Indian Institute of Science, Bangalore 560012, India\\
$^3$National Center for Radio Astrophysics, Post Bag 3, Ganeshkhind, Pune 411007, India\\
$^4$Instituto de Astrof\'isica de Andaluc\'ia, Apartado de Correos, 3.004, E-18.080, Granada, Spain \\
$^5$IRAM - Institut de Radio Astronomie Millimetrique,300 rue de la Piscine, F-38406 Saint-Martin d\' \rm Heres, France \\
$^6$Aryabhatta Research Institute of Observational Sciences, Manora Peak, Naini Tal 263129, India\\
$^7$Indian Institute of Astrophysics, Bangalore 560034, India \\
$^8$Center for Research \& Education in Science \& Technology, Hosakote, Bangalore 562114, India\\
$^9$European Southern Observatory, Alonso de C\'ordova, Casilla 19001, 
Chile \\
$^{10}$Stockholm Observatory, SE-106 91 Stockholm, Sweden\\
$^{11}$Astronomical Observatory, Nikolaev State University,
Nikolskaja, 24, Nikolaev, 54030, Ukraine \\
$^{12}$Space Telescope Science Institute, 3.700 San Mart\'{\i}n Dr,
Baltimore MD 21.218-2.463, USA}
\authorrunning{L. Resmi et al.}
\titlerunning{Radio, Millimeter and Optical Afterglow of GRB030329}
\offprints{L. Resmi, \\ \email{resmi@rri.res.in} }
\date{Received 12$^{\rm{th}}$ July 2004 / accepted  12$^{\rm{th}}$ April 2005}
\abstract{We present radio, millimeter and optical observations of the 
afterglow of GRB030329.  
$UBVR_{C}I_{C}$ photometry is presented for a period of
$3$~hours to $34$~days after the burst. Radio monitoring at 1280~MHz has
been carried out using the GMRT for more than a year. Simultaneous millimeter
observations at 90 GHz and 230 GHz have been obtained from the Swedish-ESO 
Submillimeter Telescope (SEST) and the IRAM-PdB interferometer over more than a
month following the burst. We use these data to constrain the double jet
model proposed by Berger et al. (2003) for this afterglow.  We also
examine whether instead of the two jets being simultaneously present, the
wider jet could result from the initially narrow jet, due to a fresh
supply of energy from the central engine after the ``jet break''.
}
\maketitle
\section{Introduction}
The Gamma Ray Burst of 29th March 2003 has been an unique event. At a
distance of $\sim 870$ Mpc (assuming a cosmology of $\Omega_\Lambda=0.7$
and $\Omega_m = 0.3$, and a redshift of $0.1685$ (Greiner et al. 2003a)) it
is the second nearest GRB for which an afterglow has been observed. The
optical and radio afterglow of this burst has been one of the brightest
detected till date (Peterson \& Price 2003). The spectral signature of a
supernova (SN2003dh) emerged in the optical transient a few days after the
burst (Stanek et al. 2003) and thus provided the first unambiguous
evidence of the long suspected association between Gamma Ray Bursts and
Supernovae.  Multifrequency observations indicated that the GRB consisted
of at least two jet-like components of ejection with different opening
angles and Lorentz factors, in addition to the supernova component (Berger
et al. 2003).

GRB030329 was detected and localized by the HETE-II satellite (Vanderspek
et al. 2003a).  The trigger H2652 occurred on 29th March 2003, at UT
11:37:14.7 and lasted more than $100$ s. This was one of the brightest
bursts detected by the instrument, with a $30-400$ KeV fluence of $1.1
\times 10^{-4}$ erg $\rm{cm}^{-2}$. The Soft X-ray Camera on board HETE - II
localized the burst to be at RA~($\rm{J}2000$) $= 10^{\rm{h}} 44^{\rm{m}} 49^{\rm{s}}$ and
Dec~($\rm{J}2000$) $= +21^{\circ} 28^{\prime} 44^{\prime \prime}$ within an
error circle of radius $2$~arcmin.  The temporal profile showed two
distinct peaks in the burst, separated by $\sim 11$ seconds.  Fluence in
the lower energy band, $S_{(7-30\rm{KeV})}$, was $5.5 \times 10^{-5}
\rm{erg}$ $\rm{cm}^{-2}$, which implies a hardness ratio of
$S_{(7-30\rm{KeV})}/S_{(30-400\rm{KeV})} > 0.33$, classifying this GRB
into the `X-ray rich' category (Vanderspek et al. 2004).  In a worldwide
observational campaign, the afterglow was detected in all possible
wavebands. The first X-ray detection by \textit{RXTE} $\sim 5$ hours after
the burst found the source to be extremely bright with a $2-10$ KeV flux
of $1.4 \times 10^{-10}$ erg cm$^{-2}$ sec$^{-1}$ (Marshall \& Swank 2003).  
The optical transient had an R-band magnitude $\sim 12$ when it was
reported by Peterson \& Price (2003) and Torii (2003). The VLA detected a
bright $3.5$ mJy radio afterglow at $8.46$ GHz (Berger et al. 2003) on
2003 March $30.06$ UT.  Later follow up observations in other radio
frequencies were reported by Pooley et al. (2003), Rao et al. (2003a, b),
Hoge et al. (2003) and Kuno et al. (2003). Around $7$ days after the burst
the optical spectrum showed the signature of an underlying supernova
emission (Stanek et al. 2003; Hjorth et al. 2003; Matheson et al. 2003),
and the presence of the associated supernova SN2003dh was confirmed later
by spectroscopic measurements. Continued monitoring has provided an
unprecedentedly rich temporal coverage of the transient in all wavebands
(Lipkin et al. 2004; Tiengo et al. 2003; Sheth et al. 2003; Berger et al.
2003; Guziy et al. 2005; Gorosabel et al. 2005a).

We present in this paper the observations done by an Indo-European GRB
collaboration at radio, millimeter and optical wavelengths over a total
observing span of nearly one year. The millimeter observations were
conducted for more than a month at ESO and IRAM.  Radio observations at a
frequency of 1280 MHz were carried out with the Giant Meter Wave Radio
Telescope (GMRT) (http://www.ncra.tifr.res.in) operated by the National
Center for Radio Astrophysics, Pune.  Optical follow-up was conducted till
$34$ days after the burst using the $2.01$m Himalayan Chandra Telescope
(HCT) of the Indian Astronomical Observatory (IAO), Hanle and the $1.04$m
Sampurnanand Telescope (ST) at the State Observatory, Naini Tal, now
renamed as Aryabhatta Research Institute of Observational Sciences
(ARIES).  The GMRT observations represent the lowest frequency detection
and the longest follow-up of the afterglow reported so far.  This is also
the first detection of a GRB afterglow by the GMRT. Our optical
observations in $UBVR_{C}I_{C}$ pass-bands started $\sim 3$ hours after
the burst. Except in the $R$ band, these represent the earliest photometry
of the optical transient. Our optical data also fill many temporal gaps
existing in the literature (e.g.\ Lipkin et al. 2004). At millimeter
waves, we used the SEST to make an early detection ($\sim
0.6$ days) at $86$ GHz, albeit at low statistical significance, and
subsequently monitored the burst until June 19, 2003, using the IRAM
Plateau de Bure interferometer, making several simultaneous detections at
frequencies ranging from $86$ to $240$~GHz. The early evolution of the
optical afterglow followed the familiar behavior of a power-law decay in
the light curve, with a steepening (``jet break'') around $0.5$~days
(Garnavich et al. 2003; Smith 2003; Price et al. 2003).  The nature of the
optical light curve, however, deviated from this simple model after $\sim
1.5$~days, displaying substantial variability and some change in average slope
(see Lipkin et al. 2004 for a full compilation).  At radio frequencies,
the evolution of the afterglow flux was much slower than the early optical
decay, and a steepening of the light curve was observed at $\sim 10$~days
after the burst (Berger et al. 2003).  In order to explain this behavior,
Berger et al. (2003) introduced a double-jet model, with a narrow jet
responsible for the early optical emission and a wider jet contributing to
the radio and late-time optical and X-ray emission.  In addition, the
optical light curve also has a significant contribution from the
underlying supernova SN2003dh after about a week following the burst.  
The sharp bump in the optical light curve seen at $\sim 1.5$~days has been
attributed by Berger et al. (2003) to the deceleration epoch of the wider
jet.

Granot, Nakar \& Piran (2003) have instead proposed that the bump in the
R-band lightcurve at $1.5$ days and at three successive epochs could be
attributed to refreshed shocks (see also Guziy et al. 2005). In their
original model, however, the second jet break at $\sim 10$ days was not
expected (Piran, Nakar \& Granot 2003).

In this paper, we examine the ability of the double jet model of Berger et
al.  to fit our observations along with other multi-band observations
reported in the literature.  We present a refined set of parameters for
the two jets resulting from our fits. We also examine whether the two jets
are in fact distinct entities or whether a refreshed shock might have
converted the decelerating narrow jet to a wider, more energetic jet.  In
sect. 2, 3 and 4 we describe respectively our radio, millimeter and
optical observations, in sect. 5 we present theoretical model fits and in
sect. 6 we discuss their implications. In sect. 7 we calculate the reverse
shock emission expected from the model. In sect. 8 we estimate the possible
contribution from SN2003dh.  Sec. 9 summarizes our results.

\section{Radio Observations}
We obtained radio observations at the center frequency of 1280 MHz using
the Giant Meter Wave Radio Telescope (GMRT) located at Khodad, near Pune
in Western India.  GMRT is operated by the National Center for Radio
Astrophysics. The telescope is an interferometric array of $30$ fully
steerable parabolic dishes of $45$m diameter each, spread over a distance of up to
$25$km. The telescope operates at several spot frequencies between $150$
to $1500$~MHz, with a maximum bandwidth of $32$~MHz at any given band. All
feeds provide dual polarization outputs.  The highest angular resolution
achievable ranges from about $20$ arc-sec at the lowest frequencies to
about $2$ arc-sec at $1.5$ GHz. More details about the GMRT can be found
in Swarup et al. (1991).

We made the first detection of the GRB030329 radio transient in $1280$ MHz
band on 31st March 2003, $2.3$ days after the GRB trigger (Rao et al.
2003a). The source was bright, with a flux of $0.33$ mJy, $8 \sigma$ above
the background rms noise.  Since then we have carried out a series of
monitoring observations at the center frequency of $1280$~MHz, at nine
epochs until April 18, 2004. We used a bandwidth of $32$ MHz during
these observations.  The flux scale is set by observing the primary
calibrator 3C286, 3C147 or 3C48.  A phase calibrator was observed before
and after a 30 to 45 min scan on GRB030329. The integration time was 16 s.
The data recorded from GMRT have been converted to FITS and analyzed
using Astronomical Image Processing System ({\tt AIPS}). At each epoch,
the stability of the flux scale was checked by measuring the fluxes of a
few background sources. The flux of background sources were consistently
stable at all epochs, and the rms fluctuation in their flux (estimated to
be 8\,\%) has been taken as the error in the flux calibration. The final
error quoted is the quadrature sum of the measurement error and the flux
calibration uncertainty. The rms noise in the image is estimated from a
sufficiently large area where no source was visible. The rms noise ranged
from 35 to 80 $\mu$Jy on different epochs. The radio transient was
detected in all our observations.  The flux showed an initial rise,
reaching a peak of $2.5$ mJy at $\sim 133$ days after the burst followed
by a secular decay (Table 1). As mentioned in sect. 5, the observed nature
of the 1280-MHz light curve tightly constrains the non-relativistic
transition of the expanding jet.

\section{The millimeter face of the afterglow}
The first, single dish, observations were carried out on March 29--30, 2003
at La Silla, Chile, with the 15m Swedish-ESO Submillimeter Telescope
(SEST, Booth et al. 1989).  We used the dual channel IRAM SIS-receiver,
which allows simultaneous observations at 1.3 and 3~mm wavelengths. The
1.3~mm receiver was tuned to 215.0~GHz and the 3~mm receiver was tuned to
86.243~GHz. The backend used was a 3-level correlator with 2000 channels
and the bandwidth was 1028 MHz for the 215 GHz receiver, and 512 MHz for
the 86 GHz receiver. We used a dual beam switch mode (12 arcmin throw in
azimuth) and integrated for 60 seconds (on source) per measurement.  
Observations were performed over three nights (see Table 3). During the
observations the sky was clear and stable. The pointing was checked before
and after the observations each night and was found to be stable within
$\pm$ 3$^{\prime\prime}$.

The intensity scale was calibrated using the standard chopper wheel
method, with an internal calibration error of $\sim$ 10-20$\%$. The
intensity scale is converted from K to Jy using 40 Jy/K and 25 Jy/K for
the 86 GHz and 215 GHz measurements respectively. Two blank sky
integrations were done in order to test the performance of the receiver
system (see Table 3). These observations were done during day time,
resulting in higher noise levels (16 $\pm$ 164 and 4 $\pm$ 492 mJy
respectively) than the observations of GRB030329 which were done at night
time. The blank sky observations did not produce reliable results at 215
GHz.

In order to derive the continuum level, each individual scan was inspected
for spikes (which were removed) and abnormal baseline curvature. The
average continuum and an estimate of the noise rms were derived for the
central 400 MHz of the 86 GHz spectra, and 700 MHz of the 215 GHz spectra.
The subscans were subsequently added together, weighted with their
individual noise rms. Scans deviating by more than 3 sigma from this
initial analysis were removed and the procedure repeated with the
remaining scans. This generally does not change the average continuum
flux, but decreases the dispersion. The final average continuum fluxes are
given in Table 4. The dispersion in the continuum fluxes among individual
scans are given as the 1$\sigma$ error in Table 4.

The dispersion in the measured continuum level, using the present
technique, reflects the variable sky background and the inherent
calibration uncertainty. It does not follow a normal distribution and
increasing the number of subscans does not necessarily decrease the
dispersion. The dispersion of continuum fluxes follows more a top-hat
distribution. As such, it is not a good measure of the precision of the
final continuum flux value, but nevertheless the only measure of the
uncertainty at our disposal.

Under clear and stable weather conditions, as was the case for the SEST
observations of GRB030329, the average flux value is usually a robust
estimate if the number of subscans is sufficiently large (e.g. $>$20). The
flux value derived for the night of March 29 is higher than any of the
following measurements, while the noise rms remains more or less constant.
This is a strong indication that continuum flux was detected during the
first night but not during the following nights. The weather conditions at
the higher frequency band (215 GHz) were worse than at 86 GHz. We
therefore conclude that at least at 3~mm (86 GHz), we did tentatively detect
continuum flux during March 29, at a level around 80 mJy. The uncertainty
associated with this, however, remains poorly defined and the quoted flux
value should be regarded as approximate. At a $2\sigma$ level, the 
quoted uncertainty gives an upper bound of $\sim 104$~mJy on the flux
of the source. 

At Plateau de Bure interferometer (Guilloteau et al. 1992), we carried out
simultaneous observations in two bands, around $90$~GHz and $230$~GHz, the
actual band center frequencies being slightly different on different days.
The instrument was operating on compact D configuration with six antennas
(6Dp) until May 3rd and with fewer antennas later (see Table 4). Data
reduction and analysis was performed with the GLIDAS (Grenoble Image and
Line Data Analysis) software. The target was observed at coordinates
RA~(2000) = $10^h 44^m 50.030^s$, Dec~(2000) = $+21^{\circ} 31^{\prime}
18.15^{\prime\prime}$ and the best position offsets were found
on 31 Mar 2003 to be: $-0.87^{\prime\prime} \pm  0.03^{\prime\prime}$ 
and $-0.78^{\prime\prime}\pm 0.02^{\prime\prime}$ respectively.
Therefore, the absolute coordinates are 
RA (2000) = $10^h 44^m 49.968^s (\pm 0.002 s)$, 
Dec (2000) = $+21 ^{\circ} 31 ^{\prime}$ 
$17^{\prime\prime}.37 (\pm 0.02^{\prime\prime})$.
Thus, within the error limits, the measured position definitely stays constant 
throughout the observing period. 
The source has been detected up to May 3rd, after which there are only 
upper limits available.

\section{Optical Observations and Data Reduction}
Starting about $3$~hours after the burst, we obtained a total of 
$13$, $70$, $87$, $167$, and $39$ photometric observations in Johnson $U$, 
$B$, $V$ and Cousins $R$ and $I$ bands respectively, from both ST and HCT.   
The CCD used at HCT was 1024$\times$1024 pixel$^{2}$ with the entire 
chip covering a field of $\sim 4^{\prime}.7 \times 4^{\prime}.7$ on the sky. 
It has a read out noise of 11 $e^-$ and gain of 4.8 $e^-/ADU$. 
A CCD chip of size 2048 $\times$ 2048 pixel$^{2}$ was used at ST, 
which covers a field of $\sim 13^{\prime}\times 13^{\prime}$. 
The gain and read out noise are 10 $e^-/ADU$ and 5.3 $e^-$ respectively. The frames were binned in $2 \times 2$ pixel{$^2$} to 
improve the signal-to-noise ratio of the source. 
Several twilight flat field and bias frames were obtained for the CCD images 
at both the telescopes. While imaging the optical transient (OT), several short 
exposures upto a maximum of 15 min were taken in various filters.
We used MIDAS, IRAF and DAOPHOT softwares to process the CCD frames in 
the standard fashion. The bias subtracted and flat 
fielded CCD frames were co-added, whenever found necessary. 

The $BVRI$ magnitudes of the OT obtained from the Sampurnanand Telescope at Naini Tal
were calibrated differentially using secondary standard stars no. 1, 11,
14, 19, 37 and 57 in the list of Henden (2003), while the $U$ magnitudes
were determined using reference stars 1, 11 and 37 from Henden (2003).
$BVR$ magnitudes obtained at the IAO were also calibrated
differentially, using reference stars no. 14, 18 and 19 of Henden (2003).  
OT magnitudes at similar epochs, determined using photometry from
the two sites, are consistent with each other within the limits of uncertainty.
A full compilation of our data is presented in Table~2.

Figure 1(a) displays our $UBVRI$ photometry (open circles) along with some of the data published by other authors (filled circles). As mentioned above, in $UBVI$ bands 
we report the earliest observations and in all the bands our observations 
fill in gaps in photometric data reported in the literature so far, at epochs beyond one day after the burst.

Reported X-ray observations of this afterglow are shown in an accompanying
figure 1(b). The role of the X-ray observations in constraining the model,
along with observations in other bands, is discussed in sect. 5.

\section{Modeling the Multiband Observations}
It was proposed by Rhoads (1999) that the explosion which makes the GRB may 
not be isotropic. Ejection of matter in the explosion could in fact be 
collimated within narrow cones. If so, then a 
signature of this collimation is expected in the form of an achromatic 
steepening (``jet break'') in the afterglow light curve (Rhoads 1999, 
Sari, Piran \& Halpern 1999).   The jet break appears when
the Lorentz factor of the expanding outflow drops below the inverse of the initial
angle of collimation, causing the lateral expansion of the jet to dominate 
over its radial motion. This behavior has now been observed in a number
of afterglows (see reviews by Piran 2004 and Sagar 2002).

\subsection{The double jet model for GRB030329} 
The X-ray and optical lightcurves of GRB030329 afterglow had an initial
temporal slope of $\sim -0.9$. Around half a day, the optical decay
steepened to an index of $\sim -1.9$. Sampling of the X-ray evolution was
poor, but an interpolation of the {\textit{RXTE}} and {\textit{XMM}} data
obtained during $\sim 0.1$ to $\sim 100$ days indicated a break almost at
the same time as the optical steepening mentioned above (Tiengo et al.
2003).  A nearly simultaneous break observed in frequencies separated by
four orders of magnitude suggested that the break observed at $\sim
0.5$~days was a jet break.

At radio frequencies, the first observations obtained were around $
1$~days, somewhat later than the epoch of the jet break obtained from optical and X-ray
observations. The radio and millimeter light curves, however, did not
display the behavior expected after a jet break at $0.5$~days.  Instead,
they were rather well described by a second jet, with a jet break around
10 days (Berger et al. 2003; Sheth et al. 2003).  The optical light curve
showed a re-brightening around $1.5$~days, followed by a slower decay
consistent with that expected from the second jet to which the radio
emission was attributed (Berger et al. 2003; Lipkin et al. 2004).  This
led Berger et al. (2003) to propose a double-jet model for GRB030329.  
The optical re-brightening at $1.5$~days was attributed to the epoch of
deceleration of the second jet.  Berger et al. (2003) estimated the
initial opening angle of the first (narrow) jet to be $\sim 5^{\circ}$ and
that of the second (wide) jet to be $\sim 18.4^{\circ}$. The energy
contents of the two jets were estimated to be $\sim 6.7 \times
10^{48}$~erg for the narrow jet and $\sim 10^{49}$~erg for the wide jet.  
Compared to the narrow jet, a later deceleration epoch implied a smaller
initial Lorentz factor for the wide jet, and correspondingly a much higher
initial baryon load.  The double jet model has been found to be consistent
with most observations reported till date. The optical emission at late
times ($t > 10$ days) is, however, dominated by the associated supernova
SN2003dh.

We attempted modeling our observations along with multiwavelength data
available in the literature within the ambit of this double-jet model
(model 1). The basic quantity in our model is the synchrotron source
function, which we consider to have appropriate power-law forms between
the usual break frequencies. Transition from one power-law phase to
another is made gradual through a Band type smoothening (Band et. al.
1993) at the peak frequency $\nu_m$ and the cooling frequency $\nu_c$. 
The self-absorption frequency $\nu_a$
is not treated as a break; instead the absorption is incorporated into the
expression for synchrotron optical depth, which, along with the source
function, yields the flux at any given frequency. In the co-moving frame
of the shock, the optical depth is set to unity at $\nu =\nu_a$(comoving). 
We incorporate transition to a non-relativistic phase of expansion, as in
Frail, Waxman \& Kulkarni (2000). The non-relativistic transition is
treated as a sharp break at a time $t= t_{\rm{nr}}$. We obtain our fits
through the usual $\chi^2$ minimization procedure, using $\nu_a, \nu_m, 
\nu_c, t_{\rm{nr}}$, the electron distribution index $p$ and the jet break 
time $t_j$ as the fit parameters. Here we are trying to model the
underlying smooth power law behavior rather than the short time scale
variabilities in the lightcurve. Since our model does not include the
short-term variabilities, the nominal $\chi^2$ obtained is relatively high.  
The best fit from this model is shown in figures $1$ and $2$. We derived the physical
parameters $E_{\rm{iso}}$ (the isotropic equivalent energy of the burst), $n$ (number
density of the ambient medium), $\epsilon_e$ and $\epsilon_B$ (fractional
energy content in the electrons and in the magnetic field respectively)
using the expressions in Wijers \& Galama (1999) from these fitted parameters, 
with appropriate modifications to place $\nu_a$ at
$\tau_{\nu} = 1$ instead of $0.35$ used by Wijers \& Galama (1999).

\subsubsection{The wide jet}
The parameters of the wide jet are well constrained by data in the
4--250~GHz range, as discussed by Berger et al. (2003) and Sheth et al.
(2003). At the jet break time of $9.8$~days, we find the self absorption
frequency $\nu_a$ to be $1.3^{+0.25}_{-0.06} \times 10^{10}$~Hz, the
synchrotron peak frequency $\nu_m$ to be $3.98^{+0.5}_{-0.1} \times
10^{10}$~Hz with a peak flux $F_m$ of $44.7^{+1.0}_{-2.0}$~mJy.  The post
jet-break decay in radio, the peak optical flux at $1.5$~days and the late
X-ray observations at $37$ and $61$ days together constrain the electron
energy distribution index $p$ to $2.3^{+0.05}_{-0.02}$ and the cooling
frequency $\nu_c$ to $3.98^{+1.3}_{-2.0} \times 10^{14}$~Hz after the jet
break.  These parameter values are very similar to those derived by Berger
et al. (2003).  Our $1280$~MHz observations show a gradual rise of flux to
a peak around $133$~days.  This behavior can be reproduced by a
non-relativistic transition of the jet at $t_{\rm nr}=42^{+17}_{-7}$~days.

\subsubsection{The narrow jet}
The jet break time for the narrow jet is derived to be
$0.69^{+0.08}_{-0.06}$~days from the optical light curve.  Using a galactic
extinction $E(B-V) = 0.025$ mag in the direction of the GRB (Schlegel et
al. 1998), the early optical and X-ray observations, both before and after
this break, are well described by an electron energy distribution index
$p$ of $2.12 \pm 0.05$.  Some authors (e.g.\ Sato et al. 2003, 
Lipkin et al. 2004) have conjectured that there occurs a passage of the cooling 
break ($\nu_c$) through optical bands within the first few hours after 
the GRB, based on a derived change in slope of R-band light curve around 
$\sim 0.25$~days, and a small color evolution.  Our own observations
have a continuous coverage from $\sim 0.15$~days to $\sim 0.3$~days after
the burst, taken from the same instrument and calibrated uniformly on the
same scale.  We find that our data over this interval are fit very well
by a single power law.  Even the data reported by Lipkin et al. (2004)
and Sato et al. (2003) do not conclusively demonstrate a secular steepening
at $\sim 0.25$~days; the effect could be easily mimicked by short-term
variability riding on a single, underlying power law. Colors derived 
from our own multiband observations have somewhat large errors 
($\sim 0.08$~mag $1$-$\sigma$), and we are unable to discern the $\sim 0.1$~mag 
systematic change in $B-R$ color reported by Lipkin et al. (2004).   
As mentioned by Lipkin et al. (2004), there could be various reasons
for this early color evolution.  We do not feel that the passage of
cooling break through optical bands at early times can be conclusively
established from the existing observations. From multiband fits, however, 
we estimate $\nu_c$ to be at $1.0^{+1.0}_{-0.5} \times 10^{16}$ Hz at
$0.5$~days.  At times $> 1.5$~days, the contribution from the wide jet is
sufficient to reproduce the radio light curves (see fig. 2), so the radio
emission from the narrow jet is constrained to be almost negligible.  In
order to achieve this, we need to have the peak frequency $\nu_m$ and the
self absorption frequency $\nu_a$ of the narrow jet to be as high as
possible.  The passage of $\nu_m$ is not observed through the optical
band, so we chose it to be just below the $R$-band at the earliest epoch
($\sim 0.05$~days) at which data are available.  This results in $\nu_m$
regressing to $\sim 10^{13}$~Hz at the jet break epoch of $0.5$~days. The
fitted peak flux $F_{(\nu =\nu_{m})}$ is $19.8^{+9.3}_{-2.4}$ mJy at this
time. The density of the ambient medium can be derived from the rather
well-constrained parameters of the wide jet, and works out to be $n \sim
8$ atom/cc.  This, along with the narrow jet parameters mentioned above,
predicts the self absorption frequency of the narrow jet to be $\nu_a$
equal to $3.1 ^{+0.14}_{-0.63}\times 10^{9}$ Hz at $0.5$ days.  We find
that this value of $\nu_a$ yields adequate suppression of the narrow jet
flux at low frequencies for the model to be consistent with our 1280 MHz
data.

As mentioned earlier, the $\chi^{2}_{\rm{DOF}}$ is somewhat high due to
short-term variabilities in the observed light curve. From the fit we excluded 
the first seven days
of data at $4.86$ and $8.46$~GHz, which appear to have been affected by
scintillations. The first data point at $250$~GHz ($\sim 1.5$ days) and at
$100$~GHz ($0.8$ days) were also removed from the fit. We did not consider
five more data points in radio bands (of $2.6$ days in $43$~GHz, days $1$ 
and $12$ in $22$~GHz and days $3.5$ and $4.7$ in $15$~GHz) which
produced high $\chi^2$ values due to scatter. We exclude the optical data
from discussion for epochs larger than $\sim 5$ days because of the dominant
contribution from SN2003dh. A $\chi^2_{\rm{DOF}}$ of $23.3$ is obtained
for the best fit with this model. The optical (mostly $V$ and $B$) bands
dominate the contribution to $\chi^2$ along with the lower radio
frequencies ($4$~GHz, $8$~GHz and $15$~GHz). The number density of the
ambient medium is inferred to be $8.6^{+12}_{-5}$. We infer the fractional
energy content in relativistic electrons and magnetic field to be $
0.56^{+0.4}_{-0.5}$ and $ 4^{+1.9}_{-1.8} \times 10^{-4}$ respectively for
the narrow jet, and $ 9.0^{+3}_{-1} \times 10^{-2}$ and $11.9^{+10}_{-7} \times 10^{-4}$
for the wide jet.  We derive $1.4^{+1.3}_{-0.8} \times 10^{51}$~erg for
the isotropic equivalent energy and $6.2^{+0.02}_{-0.03}$ degrees. for
the opening angle of narrow jet. This corresponds to a total energy
content of $3.3 ^{+4.8}_{-2.4} \times 10^{48}$ erg in the jet. For the
wide jet, we derive an isotopic equivalent energy of $1.2^{+0.4}_{-0.2}
\times 10^{50}$ erg, opening angle of $23.3^{+0.07}_{-0.04}$ degrees.
and a total energy of $5.0^{+3.3}_{-2.1} \times 10^{48}$ erg.

We calculated the rising flux from the wide jet at $t < 1.5$ days, assuming
time evolution of the spectral parameters to be of the form $\nu_m \propto t^0$, $\nu_c
\propto t^{-2}$, $f_{\nu_m} \propto t^3$ (Peng, K\"onigl \& Granot 2004)
and $\nu_a \propto t^{1}$ and normalizing them at $1.5$ days.

We also explored the possibility of the ambient medium of the burst being
generated by a stellar wind, with a density profile of $n(r) \propto
r^{-2}$, but were unable to obtain consistent fits with the double-jet
model.  If the model is tuned to reproduce radio data in the $8$--$43$~GHz 
frequency range, it leads to an overprediction of fluxes in
millimeter bands and an underprediction at $1280$~MHz (cf.\ Fig. 2).
\subsection{Refreshed Jet ?} Most of the observations are well reproduced
by a model which sums the contributions from the wide and narrow jet. We
note that the contribution of the narrow jet is almost negligible at radio
bands; in fact the wide jet alone is quite sufficient to account for the
observed flux after $\sim 1.5$~days.  It therefore appears to us that the
data could be well described if, instead of both jets contributing
simultaneously to the emission, the narrow jet alone contributes at epochs
earlier than $\sim 1.5$~days, and only the wide jet after that time. This
suggests that a possible re-energization event around or before $\sim
1.5$~days could have refreshed the initially narrow jet, which had entered
the lateral expansion phase, and given it additional forward momentum,
converting it into the second, `wide' jet.  The opening angle of the
laterally expanding, initially narrow jet around $\sim 1.5$~days is
estimated to be $\sim 20^{\circ}$, not far from the initial opening angle
inferred for the wide jet ($\sim 23^{\circ}$).  We therefore consider it
possible that a re-energization event occurred after the jet break of the
narrow jet as suggested by Granot, Nakar \& Piran (2003) and the
double-jet model for GRB030329 could represent the conversion of an
initially narrow jet to a wide one by the re-energization.

In a simple representation of such a re-energization, we assume that the
physical parameters of the fireball, namely $E, \epsilon_e$ and $\epsilon_B$
undergo a change after re-energization, while the ambient density $n$
remains the same. We allow these physical parameters to be determined by
the model fits.

Fits to multiwavelength observations with this model (model 2) are shown
in figs. 3 and 4. We excluded the same set of data in these fits as done in model 1. The
minimum $\chi^{2}_{\rm{DOF}}$ we obtained with this model is $24.5$,
slightly higher than the value for the previous model. Here again, the
$\chi^{2}$ is dominated by the optical band as well as the low radio
frequencies. Both model 1 and model 2 underpredict the flux at $1280$~MHz
peak by a factor of 2. But it is difficult to make a distinction between
model 1 and model 2 from the small difference in the $\chi^{2}$.
Parameters for the initial, narrow jet are nearly the same as those listed
in sect. 5.1.2.  except the self absorption frequency $\nu_a$
($2.9^{+0.8}_{-0.06} \times 10^{9}$~Hz), which is implied by the value of
$n$ inferred from the parameters of the jet after re-energization.
Parameters of the refreshed jet are only marginally different from that of
the wide jet discussed in sect. 5.1.1. We obtain a $p$ of $2.24 \pm 0.02$,
$t_j$ of $10^{+2.3}_{-1.0}$ days, and at $9.8$ days, a cooling frequency
of $5.0^{+2.1}_{-1.5} \times 10^{14}$~Hz, and a marginally reduced self
absorption frequency of $1.1^{+0.3}_{-0.05} \times 10^{10}$~Hz. The
non-relativistic transition time $t_{\rm{nr}}$ is $63^{+13.5}_{-30}$ days.
The number density of the ambient medium is inferred to be
$6.7^{+13.}_{-3.}$. We infer the fractional energy content in relativistic
electrons ($\epsilon_e$) and magnetic field ($\epsilon_B$) to be
$0.53^{+0.5}_{-0.4}$ and $4^{+2.5}_{-1.3} \times 10^{-4}$ respectively for
the initial (narrow) jet, and $0.103^{+0.05}_{-0.01}$ and $1.0^{+1}_{-0.5}
\times 10^{-3}$ for the refreshed (wide) jet.  We derive an isotropic
equivalent energy of $1.2^{+1.6}_{-0.6} \times 10^{51}$~erg for the
original jet, which has an initial opening angle of
$5.9^{+0.3}_{-0.1}$ degrees.  This corresponds to a total energy content
of $3.2^{+6}_{-2.8} \times 10^{48}$ ergs in the jet.  After
re-energization, which is assumed to end around $\sim 1.5$ days, the total
energy content of the jet increases to $5.8^{+5.9}_{-1.7} \times
10^{48}$~erg, and the jet widens to $20.5^{+0.1}_{-0.03}$ degrees.

The transition to the refreshed physical parameters of the jet is expected
to be gradual, over the time required to establish a new equilibrium.  
The timescale for achieving a new Blandford-McKee structure will be
roughly equal to the time the second wave requires to cross the existing
shocked shell. This time, in the co-moving frame, can be estimated as
the thickness of the matter in the co-moving frame of the shock, divided by
$c$, which in the observer's frame will be $\Delta t \approx R/(a
\Gamma_{\rm{old}}\Gamma_{\rm{new}}c)$, where $a \sim 5 - 10$. At $1.5$
days, when the new power law phase begins, we calculate the bulk Lorentz
factor ($\Gamma_{\rm{new}}$) to be $\sim 2.3$, by extrapolating the value
of $\Gamma$ from the jet break time. Extrapolation for the initial jet
produces $\Gamma_{\rm{old}}$ to be close to this value. Hence
$\Delta t$ will be of the order of $0.2 -0.3$ days. From a close
examination of the optical lightcurve, we find that the refreshment
episode could begin at $\sim 1.2$ days and at $\sim 1.5$ days, the new
power law phase begins.

\section{Discussion}
We have seen above a comparison between the parameters of the two models.
We now discuss their compatibility with the constraints imposed by
other available observations of
the afterglow. The angular size of the fireball estimated by Taylor et al.
(2004) can be reproduced by models with isotropic equivalent energy to
external density ratio in the range $10^{50} -10^{52}$ erg cm$^{-3}$ (Oren, Nakar \&
Piran 2004). The parameters extracted from both the models fall close to
this range.

Polarization measurements of the afterglow are available in optical
(Greiner et al. 2003b) and in radio (Taylor et al. 2004) bands. In the
optical, the degree of polarization decreases shortly after the jet break
at $\sim 0.5$ days and rapid variations in polarization start occurring
around $1.5$ days, which according to Greiner et al. could be the beginning
of a new power law phase. The linear polarization at $8$~GHz ($< 0.1\%$)
around $8$ days is significantly lower than the optical polarization
($\sim 2 \%$), which could be due to the fireball being optically thick at
this frequency. This polarization behavior has been thought to support
the double jet model (Greiner et al. 2003b; Taylor et al. 2004); however
this is equally applicable to the refreshed jet.

We point out that the first millimeter-wave observation at $86$~GHz by
the SEST at $0.6$ days and the $100$~GHz observation reported by
Sheth et al. (2003) at $0.8$ days are not well fit by this simple model.
Nor does the double-jet model of Berger et al. (2003) succeed in
reproducing this well.

\section{Reverse Shock}	
It has been pointed out (Piran, Nakar \& Granot 2003) that the
deceleration of the wide jet at $1.5$ days is expected to be accompanied
by a strong radio flash from the reverse shock, which is not observed.
This concern remains for the refreshed jet model, too. The two models differ 
in the nature of the medium in which the second shock
front decelerates (in model 1, a normal ISM while in model 2, it is the
material already shocked by the first shell). The Sedov length in these 
two cases are also to be evaluated differently, since in model 1, the
shell responsible for the wide jet encounters matter all the way from the
progenitor star while in model 2, the second wave of energy passes through
the region evacuated by the first jet.

We estimated the reverse shock emission expected in either model,
assuming that the shock is ultrarelativistic (thick shell case) (Sari \&
Piran 1995; Kobayashi 2000). For model 1 there are four relevant
regions, namely, normal ISM (0), ISM shocked by the wide jet (1), reverse shocked
ejecta (2) and the cold second shell (3). Region 1 and region 2 are
separated by a contact discontinuity (CD). We followed the formulation of
Kobayashi (2000) to obtain the flux expected from reverse shock. In model
2, since the reverse shock originates when the second wave of energy
decelerates into the already shocked material, the space ahead of the CD
will be divided into three regions instead of two (Kumar \& Piran 2000).
The five relevant regions in this case are: the ISM (0), the ISM shocked 
by the first wave (1), ISM additionally shocked by the second wave (2), reverse
shocked ejecta (3) and the cold second shell (4). A contact discontinuity
separates regions 2 and 3. Assuming pressure balance at the CD, and for the 
ejecta using the assumption that $n_4R^2$ is constant, (where $n_4$ is the 
number density of the cold shell and $R$ is
the distance to the CD), one obtains the bulk Lorentz factor ($\gamma_3$)
and thermodynamic quantities (density and pressure) of region 3.  These
quantities allow one to estimate the synchrotron emission from that region.  
The thickness ($\Delta$) of the ejecta is an unknown parameter in both 
models.  For a given $\Delta$ the peak flux produced by model 1 is two
orders of magnitude lower than model 2 at the deceleration time of
$1.5$~days. The computed flux is inversely related to $\Delta$, and the
minimum value $\Delta$ can reach without overpredicting the flux observed
at various bands is $\sim 10^{10}$ cm for model 1, and $\sim 10^{13}$~cm for model 2. 
More detailed investigations
taking into account detailed hydrodynamics in non-spherical geometry as
well as the density structure of a post-jet break fireball may be
necessary to get a better estimate.

\section{SN 2003dh}
The optical emission observed at times later than $\sim 7$~days cannot be
fully accounted for by the afterglow models discussed so far.  We
attribute the excess emission at these late epochs primarily to the
associated supernova, SN2003dh.  We subtract the afterglow flux predicted
by the model and the flux due to the host galaxy (Gorosabel et al. 2005b) from the
observed data to estimate the contribution from the supernova. Fig. 5 displays
the flux attributed to the supernova in the two models discussed above. 
The K-corrected light curve of SN1998bw (Galama et al. 1998)
appropriate for the redshift of GRB030329 is shown along with the
residuals for comparison. While being similar in temporal behavior, the residuals from
model 1 are fainter by $\sim 0.3$~mag in comparison to an equivalent
SN1998bw lightcurve. In case of model 2, this difference is $\sim
0.4$~mag. These results compare well with the estimate of Lipkin et al.
who find that an SN1998bw lightcurve, diminished by $0.3$ magnitude, is
required to fit the observed data.

\section{Summary}
We have presented low frequency radio, millimeter wave and optical
observations of the afterglow of GRB030329, and interpreted them in terms
of the double-jet model discussed earlier in the literature.  Our main
conclusions are summarized below.
\begin{itemize}
\item The $1280$ MHz GMRT observations, starting $\sim 2$ days after 
   the burst and continuing for over a year, provide constraints
   on the self absorption frequency of the emission region as well 
   as the epoch of non-relativistic transition of the fireball.
\item We report several simultaneous two-band ($\sim 90$ and $\sim 250$~GHz) 
   detections of the afterglow for over a month after the burst.
\item Our optical observations span a temporal range of $3$ hours
   to $34$ days and fill in many gaps in the coverage reported so far
   in the literature. In $UBVI$ bands our data represent the earliest 
   photometry of the optical transient.
\item We find that the data can be well fit by a model where the two
   jets are present either simultaneously or in exclusion of each other. We 
   derive a non-relativistic transition time of $\sim 42$~days for model 1 
   and $\sim 63$~days for model 2. Although concerns with reverse shock 
   emission remain in both models, we deem it possible that the optical 
   re-brightening seen at the epoch of $\sim 1.3$~days could be a 
   re-energization of the jet, which resulted in the initially narrow jet 
   being converted into a more energetic wider jet.
\end{itemize}

\section*{Acknowledgments}
The GMRT is operated by the National Center For Radio Astrophysics of the
Tata Institute of Fundamental Research. This work is based partly on
observations carried out with the IRAM Plateau de Bure Interferometer.
IRAM is supported by INSU/CNRS (France), MPG (Germany) and IGN(Spain). The
SEST observations were conducted as part of the GRACE Collaboration
Program for the multiwavelength follow up of Gamma-ray Bursts at ESO
with ID 70.D-0523. We thank the ESO and IRAM Plateau de Bure staff and
their operators for their great support in the observations. The HCT is
operated by the Indian Institute of Astrophysics, Bangalore. This paper
makes use of data obtained through the High Energy Astrophysics Science
Archive Research Center Online Service, provided by the NASA/Goddard Space
Flight Center. The GCN system, managed and operated by Scott Barthelmy, is
gratefully acknowledged. This research is partially supported by the
Spain's Ministerio de Educaci\'on y Ciencia through programs
ESP2002-04124-C03-01 and AYA2004-01515 (including FEDER funds). LR
acknowledges financial support from Council for Scientific and Industrial
Research, India.  The authors thank an anonymous referee for critical
comments which helped improve the paper.
%
%------------------------------REFERENCES----------------------
%

%
%-------------------------Table 1: GMRT data--------------------------
%
\begin{table}
\begin{center}
 {\bf Table 1.}~$1280$ MHz radio Observations at GMRT \\
\begin{tabular}{ccll}\hline
&Mid UT & Flux (mJy) \\ \hline
2003&& \\
& Mar 31 18:30   & 0.33 +/- 0.09\\ 
&Apr 1 20:30  & 0.33 +/- 0.09\\ 
&May 30 16:30  &   1.10  +/- 0.15  \\
&Jun 28 11:00  &   1.40  +/- 0.16  \\
&Aug 9 07:30   &    2.50  +/- 0.25  \\
&Oct 20 05:00  &    1.50  +/- 0.17  \\
&Dec 3 00:20   &    1.30  +/- 0.14  \\
2004&& \\
& Jan 29 22:12  &   1.20 +/- 0.18     \\
&Apr 18 14:32   &   1.10 +/- 0.12  \\ \hline
\end{tabular}
\end{center}
\end{table}
%
%--------------------------Table 2 : Optical Data----------------
% U band starts here
\begin{table}
\begin{center}
 {\bf Table 2.}~Optical Transient \\
\begin{tabular}{ccccll} \hline 
&Date (UT) &Time since Burst&Magnitude & Exposure time & Telescope  \\
         &&in days &  (mag)&(Seconds)&  \\   \hline 
\multicolumn{5}{c}{\bf $U-$ passband}  \\
2003 & March&&&&\\
& 29.6459 & 0.1619 & 13.606 $\pm$ 0.051 & 200 & ST \\
& 29.6918 & 0.2078 & 13.824 $\pm$ 0.052 & 200 & ST \\
& 29.7042 & 0.2202 & 13.831 $\pm$ 0.060 & 200 & ST \\
& 29.7186 & 0.2346 & 13.910 $\pm$ 0.061 & 200 & ST \\
& 29.7345 & 0.2505 & 14.016 $\pm$ 0.053 & 200 & ST \\
& 29.7477 & 0.2637 & 14.069 $\pm$ 0.051 & 200 & ST \\
& 30.7558 & 1.2718 & 16.454 $\pm$ 0.053 & 300 & ST \\
& 30.7738 & 1.2898 & 16.486 $\pm$ 0.057 & 300 & ST \\
& 30.7909 & 1.3069 & 16.483 $\pm$ 0.056 & 300 & ST \\
& 30.8084 & 1.3244 & 16.507 $\pm$ 0.063 & 300 & ST \\
2003 & April&&&&\\
& 1.6524 & 3.1684 & 17.156 $\pm$ 0.061 & 400 & ST \\
& 1.7046 & 3.2206 & 17.153 $\pm$ 0.058 & 400 & ST \\
& 1.7626 & 3.2786 & 17.201 $\pm$ 0.055 & 400 & ST \\
\hline
%
%  B band starts here
%
\multicolumn{5}{c}{\bf $B-$ passband}  \\
2003 & March&&&&\\
& 29.6429 & 0.1589 & 14.189 $\pm$ 0.026 & 150 & ST \\
& 29.6483 & 0.1643 & 14.1463 $\pm$ 0.070 & 900 & HCT \\
& 29.6552 & 0.1712 & 14.248 $\pm$ 0.032 & 100 & ST \\
& 29.6603 & 0.1763 & 14.1933 $\pm$ 0.046 & 900 & HCT \\
& 29.6891 & 0.2051 & 14.409 $\pm$ 0.031 & 100 & ST \\
& 29.7013 & 0.2173 & 14.440 $\pm$ 0.022 & 100 & ST \\
& 29.7153 & 0.2313 & 14.534 $\pm$ 0.012 & 200 & ST \\
& 29.7312 & 0.2472 & 14.599 $\pm$ 0.016 & 200 & ST \\
& 29.7445 & 0.2605 & 14.653 $\pm$ 0.014 & 200 & ST \\
& 29.7574 & 0.2734 & 14.704 $\pm$ 0.028 & 200 & ST \\
& 30.7378 & 1.2538 & 17.0433 $\pm$ 0.033 & 900 & HCT \\
& 30.7598 & 1.2758 & 17.137 $\pm$ 0.015 & 200 & ST \\
& 30.7730 & 1.289 & 17.0573 $\pm$ 0.060 & 900 & HCT \\
& 30.7776 & 1.2936 & 17.170 $\pm$ 0.016 & 200 & ST \\
& 30.7948 & 1.3108 & 17.167 $\pm$ 0.014 & 200 & ST \\
& 30.8123 & 1.3283 & 17.192 $\pm$ 0.016 & 200 & ST \\
& 30.8359 & 1.3519 & 17.0923 $\pm$ 0.032 & 900 & HCT \\
& 31.5978 & 2.1138 & 17.464 $\pm$ 0.032 & 300 & ST \\
& 31.6707 & 2.1867 & 17.3683 $\pm$ 0.053 & 900 & HCT \\
& 31.6763 & 2.1923 & 17.421 $\pm$ 0.032 & 300 & ST \\
& 31.6984 & 2.2144 & 17.472 $\pm$ 0.030 & 500 & ST \\
& 31.7192 & 2.2352 & 17.496 $\pm$ 0.031 & 300 & ST \\
\hline
\end{tabular}
\end{center}
\end{table}
%%%%           split in B band
\begin{table}
\begin{center}
 {\bf Table 2.}~Optical Transient (contd.)\\
\begin{tabular}{ccccll} \hline 
& 31.7410 & 2.257 & 17.4503 $\pm$ 0.028 & 900 & HCT \\
& 31.7414 & 2.2574 & 17.509 $\pm$ 0.030 & 300 & ST \\
& 31.7598 & 2.2758 & 17.508 $\pm$ 0.029 & 300 & ST \\
& 31.7740 & 2.29 & 17.4403 $\pm$ 0.029 & 900 & HCT \\
& 31.7769 & 2.2929 & 17.485 $\pm$ 0.031 & 300 & ST \\
& 31.7970 & 2.313 & 17.520 $\pm$ 0.029 & 300 & ST \\
& 31.8046 & 2.3206 & 17.4553 $\pm$ 0.020 & 900 & HCT \\
& 31.8149 & 2.3309 & 17.550 $\pm$ 0.028 & 400 & ST \\
& 31.8343 & 2.3503 & 17.4853 $\pm$ 0.028 & 900 & HCT \\
& 31.8387 & 2.3547 & 17.586 $\pm$ 0.029 & 400 & ST \\
& 31.8600 & 2.376 & 17.595 $\pm$ 0.031 & 400 & ST \\
& 31.8806 & 2.3966 & 17.617 $\pm$ 0.022 & 400 & ST \\
2003&April&&&&\\
& 1.5967 & 3.1127 & 17.866 $\pm$ 0.033 & 300 & ST \\
& 1.6277 & 3.1437 & 17.749 $\pm$ 0.038 & 200 & ST \\
& 1.7153 & 3.2313 & 17.826 $\pm$ 0.029 & 400 & ST \\
& 1.7207 & 3.2367 & 17.831 $\pm$ 0.027 & 400 & ST \\
& 1.7356 & 3.2516 & 17.7193 $\pm$ 0.031 & 900 & HCT \\
& 1.8582 & 3.3742 & 17.791 $\pm$ 0.037 & 300 & ST \\
& 1.9000 & 3.416 & 17.7843 $\pm$ 0.034 & 900 & HCT \\
& 2.6386 & 4.1546 & 18.3483 $\pm$ 0.036 & 900 & HCT \\
& 2.6436 & 4.1596 & 18.419 $\pm$ 0.041 & 300 & ST \\
& 2.6771 & 4.1931 & 18.4003 $\pm$ 0.029 & 900 & HCT \\
& 2.6862 & 4.2022 & 18.534 $\pm$ 0.039 & 300 & ST \\
& 2.7439 & 4.2599 & 18.528 $\pm$ 0.038 & 300 & ST \\
& 2.7707 & 4.2867 & 18.506 $\pm$ 0.039 & 300 & ST \\
& 4.6673 & 6.1833 & 18.787 $\pm$ 0.039 & 400 & ST \\
& 4.6966 & 6.2126 & 18.748 $\pm$ 0.034 & 400 & ST \\
& 4.7246 & 6.2406 & 18.852 $\pm$ 0.033 & 400 & ST \\
& 5.6125 & 7.1285 & 19.2573 $\pm$ 0.067 & 900 & HCT \\
& 5.6511 & 7.1671 & 19.2373 $\pm$ 0.053 & 900 & HCT \\
& 5.7416 & 7.2576 & 19.2963 $\pm$ 0.046 & 900 & HCT \\
& 5.7702 & 7.2862 & 19.3183 $\pm$ 0.044 & 900 & HCT \\
& 5.8037 & 7.3197 & 19.3153 $\pm$ 0.046 & 900 & HCT \\
& 5.8293 & 7.3453 & 19.3683 $\pm$ 0.030 & 900 & HCT \\
& 5.8618 & 7.3778 & 19.3873 $\pm$ 0.045 & 900 & HCT \\
& 6.6438 & 8.1598 & 19.6213 $\pm$ 0.027 & 900 & HCT \\
& 6.6559 & 8.1719 & 19.6443 $\pm$ 0.028 & 900 & HCT \\
& 6.7522 & 8.2682 & 19.6063 $\pm$ 0.030 & 900 & HCT \\
& 6.8384 & 8.3544 & 19.6183 $\pm$ 0.039 & 900 & HCT \\
& 6.8505 & 8.3665 & 19.6553 $\pm$ 0.034 & 900 & HCT \\
\hline
\end{tabular}
\end{center}
\end{table}
%%%                  split in B band
\begin{table}
\begin{center}
 {\bf Table 2.}~Optical Transient (contd.)\\
\begin{tabular}{ccccll} \hline 
& 7.6122 & 9.1282 & 19.761 $\pm$ 0.144 & 400 & ST \\
& 8.6260 & 10.142 & 19.983 $\pm$ 0.069 & 600 & ST \\
& 8.6641 & 10.1801 & 19.943 $\pm$ 0.076 & 600 & ST \\
& 8.6933 & 10.2093 & 19.919 $\pm$ 0.064 & 600 & ST \\
& 10.6368 & 12.1528 & 20.287 $\pm$ 0.087 & 900 & ST \\
& 10.7059 & 12.2219 & 20.1933 $\pm$ 0.030 & 600X4 & HCT \\
& 11.6782 & 13.1942 & 20.432 $\pm$ 0.177 & 600 & ST \\
2003&May&&&&\\
& 1.6271 & 33.1431 & 22.214 $\pm$ 0.156 & 3*500+900 & ST \\
\hline
%
%              V band starts here
%
\multicolumn{5}{c}{\bf $V-$ passband}  \\
2003 & March&&&&\\
& 29.6310 & 0.147& 13.8983 $\pm$ 0.029 & 600 & HCT \\
& 29.6363 & 0.1523& 13.848 $\pm$ 0.020 & 200 & ST \\
& 29.6515 & 0.1675& 13.930 $\pm$ 0.023 & 75 & ST \\
& 29.6872 & 0.2032& 14.087 $\pm$ 0.021 & 50 & ST \\
& 29.6991 & 0.2151& 14.142 $\pm$ 0.017 & 100 & ST \\
& 29.7127 & 0.2287& 14.213 $\pm$ 0.018 & 100 & ST \\
& 29.7284 & 0.2444& 14.266 $\pm$ 0.033 & 100 & ST \\
& 29.7416 & 0.2576& 14.331 $\pm$ 0.018 & 100 & ST \\
& 29.7552 & 0.2712& 14.371 $\pm$ 0.032 & 100 & ST \\
& 29.7985 & 0.3145& 14.569 $\pm$ 0.032 & 200 & ST \\
& 29.8773 & 0.3933& 14.9323 $\pm$ 0.050 & 600 & HCT \\
& 30.6674 & 1.1834& 16.7423 $\pm$ 0.047 & 600 & HCT \\
& 30.7486 & 1.2646& 16.7943 $\pm$ 0.050 & 600 & HCT \\
& 30.7632 & 1.2792& 16.790 $\pm$ 0.013 & 200 & ST \\
& 30.7809 & 1.2969& 16.789 $\pm$ 0.013 & 200 & ST \\
& 30.7836 & 1.2996& 16.8153 $\pm$ 0.032 & 600 & HCT \\
& 30.7980 & 1.314& 16.773 $\pm$ 0.013 & 200 & ST \\
& 30.8202 & 1.3362& 16.825 $\pm$ 0.013 & 200 & ST \\
& 30.8256 & 1.3416& 16.8363 $\pm$ 0.040 & 600 & HCT \\
& 30.8680 & 1.384& 16.8183 $\pm$ 0.046 & 600 & HCT \\
& 30.8998 & 1.4158& 16.8353 $\pm$ 0.050 & 600 & HCT \\
& 31.6026 & 2.1186& 17.111 $\pm$ 0.013 & 300 & ST \\
& 31.6531 & 2.1691& 17.1033 $\pm$ 0.064 & 600 & HCT \\
& 31.6811 & 2.1971& 17.094 $\pm$ 0.024 & 300 & ST \\
& 31.7041 & 2.2201& 17.126 $\pm$ 0.018 & 300 & ST \\
& 31.7263 & 2.2423& 17.148 $\pm$ 0.017 & 300 & ST \\
& 31.7308 & 2.2468& 17.1783 $\pm$ 0.055 & 600 & HCT \\
& 31.7459 & 2.2619& 17.150 $\pm$ 0.016 & 300 & ST \\
& 31.7589 & 2.2749& 17.1653 $\pm$ 0.052 & 600 & HCT \\
\hline
\end{tabular}
\end{center}
\end{table}
%%%                   the split in V band
\begin{table}
\begin{center}
 {\bf Table 2.}~Optical Transient (contd.)\\
\begin{tabular}{ccccll} \hline 
&Date (UT) &Time since Burst& Magnitude & Exposure time & Telescope  \\
         && in days &  (mag)&(Seconds)&  \\   \hline 
& 31.7643 & 2.2803& 17.148 $\pm$ 0.018 & 300 & ST \\
& 31.7814 & 2.2974& 17.157 $\pm$ 0.020 & 300 & ST \\
& 31.7894 & 2.3054& 17.1883 $\pm$ 0.040 & 600 & HCT \\
& 31.8015 & 2.3175& 17.179 $\pm$ 0.014 & 300 & ST \\
& 31.8211 & 2.3371& 17.212 $\pm$ 0.014 & 400 & ST \\
& 31.8216 & 2.3376& 17.2143 $\pm$ 0.062 & 600 & HCT \\
& 31.8444 & 2.3604& 17.227 $\pm$ 0.016 & 400 & ST \\
& 31.8498 & 2.3658& 17.2443 $\pm$ 0.069 & 600 & HCT \\
& 31.8656 & 2.3816& 17.253 $\pm$ 0.014 & 400 & ST \\
& 31.8781 & 2.3941& 17.2633 $\pm$ 0.055 & 600 & HCT \\
& 31.8862 & 2.4022& 17.253 $\pm$ 0.020 & 400 & ST \\
& 31.9177 & 2.4337& 17.2843 $\pm$ 0.059 & 600 & HCT \\
2003 & April&&&&\\
& 1.6139 & 3.1299 & 17.438 $\pm$ 0.017 & 300 & ST \\
& 1.6727 & 3.1887 & 17.453 $\pm$ 0.017 & 300 & ST \\
& 1.6758 & 3.1918 & 17.4463 $\pm$ 0.051 & 450 & HCT \\
& 1.7215 & 3.2375 & 17.4203 $\pm$ 0.051 & 450 & HCT \\
& 1.7326 & 3.2486 & 17.453 $\pm$ 0.016 & 400 & ST \\
& 1.7381 & 3.2541 & 17.464 $\pm$ 0.016 & 400 & ST \\
& 1.8841 & 3.4001 & 17.427 $\pm$ 0.018 & 300 & ST \\
& 1.8847 & 3.4007 & 17.4233 $\pm$ 0.040 & 450 & HCT \\
& 2.6234 & 4.1394 & 18.0493 $\pm$ 0.061 & 450 & HCT \\
& 2.6595 & 4.1755 & 18.048 $\pm$ 0.021 & 300 & ST \\
& 2.6612 & 4.1772 & 18.0873 $\pm$ 0.065 & 450 & HCT \\
& 2.6903 & 4.2063 & 18.0913 $\pm$ 0.064 & 450 & HCT \\
& 2.6909 & 4.2069 & 18.068 $\pm$ 0.020 & 300 & ST \\
& 2.7221 & 4.2381 & 18.112 $\pm$ 0.024 & 300 & ST \\
& 4.6767 & 6.1927 & 18.370 $\pm$ 0.024 & 400 & ST \\
& 4.7069 & 6.2229 & 18.393 $\pm$ 0.025 & 400 & ST \\
& 4.7302 & 6.2462 & 18.481 $\pm$ 0.023 & 400 & ST \\
& 5.5977 & 7.1137 & 18.8733 $\pm$ 0.036 & 450 & HCT \\
& 5.6670 & 7.183 & 18.8953 $\pm$ 0.039 & 450 & HCT \\
& 5.6997 & 7.2157 & 18.9383 $\pm$ 0.052 & 450 & HCT \\
& 5.7303 & 7.2463 & 18.9283 $\pm$ 0.055 & 450 & HCT \\
& 5.7573 & 7.2733 & 18.9353 $\pm$ 0.044 & 450 & HCT \\
& 6.6149 & 8.1309 & 19.1813 $\pm$ 0.027 & 600X2 & HCT \\
& 6.7218 & 8.2378 & 19.2393 $\pm$ 0.027 & 600X3 & HCT \\
& 6.8111 & 8.3271 & 19.2183 $\pm$ 0.046 & 600X3 & HCT \\
& 6.8767 & 8.3927 & 19.2253 $\pm$ 0.064 & 600X2 & HCT \\
& 7.6073 & 9.1233 & 19.294 $\pm$ 0.073 & 300 & ST \\
\hline
\end{tabular}
\end{center}
\end{table}
%%%                  split in V band %%%%
\begin{table}
\begin{center}
 {\bf Table 2.}~Optical Transient (contd.)\\
\begin{tabular}{ccccll} \hline 
&Date (UT) & Time since Burst& Magnitude & Exposure time & Telescope  \\
         & &in days&  (mag)&(Seconds)&  \\   \hline 
& 8.6170 & 10.133 & 19.364 $\pm$ 0.033 & 600 & ST \\
& 8.6567 & 10.1727 & 19.406 $\pm$ 0.037 & 600 & ST \\
& 8.6856 & 10.2016 & 19.235 $\pm$ 0.035 & 600 & ST \\
& 9.5952 & 11.1112 & 19.414 $\pm$ 0.070 & 400 & ST \\
& 9.6050 & 11.121 & 19.5083 $\pm$ 0.032 & 900 & HCT \\
& 9.6439 & 11.1599 & 19.5233 $\pm$ 0.040 & 900 & HCT \\
& 9.6730 & 11.189 & 19.5233 $\pm$ 0.035 & 300X2 & HCT \\
& 9.7002 & 11.2162 & 19.5283 $\pm$ 0.044 & 700 & HCT \\
& 9.8625 & 11.3785 & 19.5463 $\pm$ 0.046 & 900 & HCT \\
& 10.6424 & 12.1584 & 19.585 $\pm$ 0.067 & 900 & ST \\
& 10.6461 & 12.1621 & 19.6303 $\pm$ 0.036 & 900X2 & HCT \\
& 10.6656 & 12.1816 & 19.6513 $\pm$ 0.058 & 450X3 & HCT \\
& 10.7440 & 12.26 & 19.6313 $\pm$ 0.060 & 900 & HCT \\
& 11.6590 & 13.175 & 19.7303 $\pm$ 0.043 & 450X2 & HCT \\
& 11.6925 & 13.2085 & 19.7763 $\pm$ 0.039 & 450X3 & HCT \\
& 11.6975 & 13.2135 & 19.951 $\pm$ 0.174 & 600 & ST \\
& 22.6520 & 24.168 & 20.5103 $\pm$ 0.048 & 600X3 & HCT \\
& 23.6646 & 25.1806 & 20.6343 $\pm$ 0.050 & 300X2+600 & HCT \\
2003 & May&&&&\\
& 1.6383 & 33.1543 & 20.836 $\pm$ 0.123 & 900 & ST \\
\hline
%%%                  R band starts here %%%%
\multicolumn{5}{c}{\bf $R-$ passband}  \\
2003 & March&&&&\\
& 29.6133 & 0.1293 & 13.3803 $\pm$ 0.031 & 120 & HCT \\
& 29.6171 & 0.1331 & 13.3743 $\pm$ 0.041 & 300 & HCT \\
& 29.6227 & 0.1387 & 13.431 $\pm$ 0.032 & 100 & ST \\
& 29.6232 & 0.1392 & 13.4123 $\pm$ 0.060 & 450 & HCT \\
& 29.6264 & 0.1424 & 13.422 $\pm$ 0.025 & 200 & ST \\
& 29.6397 & 0.1557 & 13.506 $\pm$ 0.024 & 200 & ST \\
& 29.6483 & 0.1643 & 13.577 $\pm$ 0.024 & 50 & ST \\
& 29.6530 & 0.169 & 13.619 $\pm$ 0.034 & 50 & ST \\
& 29.6572 & 0.1732 & 13.636 $\pm$ 0.040 & 100 & ST \\
& 29.6701 & 0.1861 & 13.7243 $\pm$ 0.034 & 600 & HCT \\
& 29.6784 & 0.1944 & 13.7233 $\pm$ 0.044 & 600 & HCT \\
& 29.6839 & 0.1999 & 13.724 $\pm$ 0.025 & 50 & ST \\
& 29.6863 & 0.2023 & 13.7893 $\pm$ 0.031 & 600 & HCT \\
& 29.6967 & 0.2127 & 13.829 $\pm$ 0.026 & 100 & ST \\
& 29.7103 & 0.2263 & 13.854 $\pm$ 0.026 & 100 & ST \\
\hline
\end{tabular}
\end{center}
\end{table}
%%%                  split in R band %%%%
\begin{table}
\begin{center}
 {\bf Table 2.}~Optical Transient (contd.)\\
\begin{tabular}{ccccll} \hline 
&Date (UT) &Time since Burst& Magnitude & Exposure time & Telescope  \\
         &&in days&   (mag)&(Seconds)&  \\   \hline 
& 29.7262 & 0.2422 & 13.916 $\pm$ 0.026 & 100 & ST \\
& 29.7396 & 0.2556 & 13.982 $\pm$ 0.027 & 100 & ST \\
& 29.7526 & 0.2686 & 14.030 $\pm$ 0.027 & 100 & ST \\
& 29.7650 & 0.281 & 14.098 $\pm$ 0.052 & 200 & ST \\
& 29.7937 & 0.3097 & 14.210 $\pm$ 0.032 & 100 & ST \\
& 29.8675 & 0.3835 & 14.4893 $\pm$ 0.050 & 600 & HCT \\
& 29.8857 & 0.4017 & 14.5483 $\pm$ 0.046 & 600 & HCT \\
& 29.9079 & 0.4239 & 14.6333 $\pm$ 0.032 & 450 & HCT \\
& 29.9140 & 0.43 & 14.6533 $\pm$ 0.033 & 450 & HCT \\
& 29.9201 & 0.4361 & 14.6613 $\pm$ 0.040 & 450 & HCT \\
& 29.9275 & 0.4435 & 14.6753 $\pm$ 0.070 & 450 & HCT \\
& 30.5833 & 1.0993 & 16.249 $\pm$ 0.070 & 100 & ST \\
& 30.5875 & 1.1035 & 16.295 $\pm$ 0.056 & 200 & ST \\
& 30.6518 & 1.1678 & 16.2963 $\pm$ 0.070 & 180 & HCT \\
& 30.6586 & 1.1746 & 16.3343 $\pm$ 0.042 & 600 & HCT \\
& 30.6747 & 1.1907 & 16.3533 $\pm$ 0.039 & 450 & HCT \\
& 30.6879 & 1.2039 & 16.3733 $\pm$ 0.028 & 450 & HCT \\
& 30.7268 & 1.2428 & 16.3723 $\pm$ 0.049 & 450 & HCT \\
& 30.7637 & 1.2797 & 16.4133 $\pm$ 0.040 & 450 & HCT \\
& 30.7665 & 1.2825 & 16.420 $\pm$ 0.021 & 200 & ST \\
& 30.7841 & 1.3001 & 16.438 $\pm$ 0.018 & 200 & ST \\
& 30.7971 & 1.3131 & 16.4223 $\pm$ 0.050 & 450 & HCT \\
& 30.8012 & 1.3172 & 16.439 $\pm$ 0.018 & 200 & ST \\
& 30.8176 & 1.3336 & 16.4453 $\pm$ 0.041 & 450 & HCT \\
& 30.8238 & 1.3398 & 16.460 $\pm$ 0.017 & 200 & ST \\
& 30.8457 & 1.3617 & 16.4593 $\pm$ 0.039 & 450 & HCT \\
& 30.8603 & 1.3763 & 16.4683 $\pm$ 0.034 & 450 & HCT \\
& 30.8759 & 1.3919 & 16.4623 $\pm$ 0.032 & 450 & HCT \\
& 30.8914 & 1.4074 & 16.4613 $\pm$ 0.039 & 450 & HCT \\
& 30.9076 & 1.4236 & 16.4353 $\pm$ 0.057 & 450 & HCT \\
& 30.9142 & 1.4302 & 16.4703 $\pm$ 0.034 & 450 & HCT \\
& 30.9212 & 1.4372 & 16.4303 $\pm$ 0.031 & 450 & HCT \\
& 30.9276 & 1.4436 & 16.4013 $\pm$ 0.063 & 450 & HCT \\
& 31.6449 & 2.1609 & 16.7163 $\pm$ 0.040 & 450 & HCT \\
& 31.6610 & 2.177 & 16.7163 $\pm$ 0.031 & 450 & HCT \\
& 31.6653 & 2.1813 & 16.711 $\pm$ 0.032 & 100 & ST \\
& 31.6804 & 2.1964 & 16.7103 $\pm$ 0.041 & 450 & HCT \\
& 31.6898 & 2.2058 & 16.744 $\pm$ 0.033 & 200 & ST \\
\hline
\end{tabular}
\end{center}
\end{table}
%%%                  split in V band %%%%
\begin{table}
\begin{center}
 {\bf Table 2.}~Optical Transient (contd.)\\
\begin{tabular}{ccccll} \hline 
&Date (UT) &Time since Burst& Magnitude & Exposure time & Telescope  \\
         & &in days&  (mag)&(Seconds)&  \\   \hline 
& 31.7083 & 2.2243 & 16.771 $\pm$ 0.031 & 200 & ST \\
& 31.7148 & 2.2308 & 16.783 $\pm$ 0.027 & 300 & ST \\
& 31.7231 & 2.2391 & 16.7823 $\pm$ 0.026 & 450 & HCT \\
& 31.7303 & 2.2463 & 16.772 $\pm$ 0.027 & 200 & ST \\
& 31.7500 & 2.266 & 16.779 $\pm$ 0.026 & 200 & ST \\
& 31.7507 & 2.2667 & 16.7953 $\pm$ 0.040 & 450 & HCT \\
& 31.7655 & 2.2815 & 16.7933 $\pm$ 0.035 & 300 & HCT \\
& 31.7682 & 2.2842 & 16.788 $\pm$ 0.029 & 200 & ST \\
& 31.7826 & 2.2986 & 16.7953 $\pm$ 0.043 & 300 & HCT \\
& 31.7853 & 2.3013 & 16.798 $\pm$ 0.029 & 200 & ST \\
& 31.7943 & 2.3103 & 16.8203 $\pm$ 0.025 & 300 & HCT \\
& 31.8054 & 2.3214 & 16.820 $\pm$ 0.025 & 200 & ST \\
& 31.8139 & 2.3299 & 16.8083 $\pm$ 0.034 & 300 & HCT \\
& 31.8264 & 2.3424 & 16.8443 $\pm$ 0.043 & 300 & HCT \\
& 31.8280 & 2.344 & 16.844 $\pm$ 0.024 & 300 & ST \\
& 31.8431 & 2.3591 & 16.8503 $\pm$ 0.040 & 300 & HCT \\
& 31.8495 & 2.3655 & 16.867 $\pm$ 0.024 & 300 & ST \\
& 31.8545 & 2.3705 & 16.8583 $\pm$ 0.048 & 300 & HCT \\
& 31.8705 & 2.3865 & 16.874 $\pm$ 0.024 & 300 & ST \\
& 31.8713 & 2.3873 & 16.8573 $\pm$ 0.044 & 300 & HCT \\
& 31.8832 & 2.3992 & 16.8683 $\pm$ 0.045 & 300 & HCT \\
& 31.8912 & 2.4072 & 16.900 $\pm$ 0.023 & 300 & ST \\
& 31.8999 & 2.4159 & 16.8723 $\pm$ 0.059 & 300 & HCT \\
& 31.9111 & 2.4271 & 16.9013 $\pm$ 0.037 & 300 & HCT \\
& 31.9223 & 2.4383 & 16.9163 $\pm$ 0.028 & 300 & HCT \\
2003 & April&&&&\\
& 1.6008 & 3.1168 & 17.071 $\pm$ 0.027 & 200 & ST \\
& 1.6237 & 3.1397 & 17.045 $\pm$ 0.025 & 300 & ST \\
& 1.6569 & 3.1729 & 17.059 $\pm$ 0.025 & 200 & ST \\
& 1.6694 & 3.1854 & 17.0553 $\pm$ 0.046 & 300 & HCT \\
& 1.6766 & 3.1926 & 17.075 $\pm$ 0.026 & 200 & ST \\
& 1.6821 & 3.1981 & 17.0453 $\pm$ 0.045 & 300 & HCT \\
& 1.6950 & 3.211 & 17.079 $\pm$ 0.026 & 200 & ST \\
& 1.7274 & 3.2434 & 17.0473 $\pm$ 0.037 & 300 & HCT \\
& 1.7445 & 3.2605 & 17.0583 $\pm$ 0.032 & 300 & HCT \\
& 1.7487 & 3.2647 & 17.075 $\pm$ 0.029 & 300 & ST \\
& 1.7531 & 3.2691 & 17.071 $\pm$ 0.026 & 300 & ST \\
& 1.8628 & 3.3788 & 17.058 $\pm$ 0.023 & 300 & ST \\
& 1.8891 & 3.4051 & 17.039 $\pm$ 0.024 & 300 & ST \\
& 1.8906 & 3.4066 & 17.0273 $\pm$ 0.035 & 300 & HCT \\
\hline
\end{tabular}
\end{center}
\end{table}
%%%                  split in V band %%%%
\begin{table}
\begin{center}
 {\bf Table 2.}~Optical Transient (contd.)\\
\begin{tabular}{ccccll} \hline 
&Date (UT) &Time since Burst& Magnitude & Exposure time & Telescope  \\
         &&in days &  (mag)&(Seconds)&  \\   \hline 
& 1.9103 & 3.4263 & 17.101 $\pm$ 0.028 & 300 & ST \\
& 2.6158 & 4.1318 & 17.6723 $\pm$ 0.046 & 300 & HCT \\
& 2.6391 & 4.1551 & 17.694 $\pm$ 0.028 & 300 & ST \\
& 2.6425 & 4.1585 & 17.6933 $\pm$ 0.029 & 300 & HCT \\
& 2.6531 & 4.1691 & 17.7113 $\pm$ 0.029 & 300 & HCT \\
& 2.6553 & 4.1713 & 17.707 $\pm$ 0.025 & 300 & ST \\
& 2.6697 & 4.1857 & 17.680 $\pm$ 0.027 & 300 & ST \\
& 2.6843 & 4.2003 & 17.7193 $\pm$ 0.033 & 300 & HCT \\
& 2.6953 & 4.2113 & 17.689 $\pm$ 0.028 & 300 & ST \\
& 2.6962 & 4.2122 & 17.7023 $\pm$ 0.033 & 300 & HCT \\
& 2.7124 & 4.2284 & 17.706 $\pm$ 0.034 & 300 & ST \\
& 2.7265 & 4.2425 & 17.709 $\pm$ 0.030 & 300 & ST \\
& 2.7484 & 4.2644 & 17.721 $\pm$ 0.030 & 300 & ST \\
& 2.7615 & 4.2775 & 17.737 $\pm$ 0.030 & 300 & ST \\
& 2.7752 & 4.2912 & 17.718 $\pm$ 0.032 & 300 & ST \\
& 4.6831 & 6.1991 & 18.030 $\pm$ 0.031 & 400 & ST \\
& 4.7125 & 6.2285 & 18.085 $\pm$ 0.035 & 400 & ST \\
& 4.7357 & 6.2517 & 18.130 $\pm$ 0.031 & 400 & ST \\
& 5.5919 & 7.1079 & 18.4583 $\pm$ 0.049 & 300 & HCT \\
& 5.6044 & 7.1204 & 18.5093 $\pm$ 0.028 & 300 & HCT \\
& 5.6209 & 7.1369 & 18.5033 $\pm$ 0.025 & 300 & HCT \\
& 5.6340 & 7.15 & 18.4963 $\pm$ 0.032 & 300 & HCT \\
& 5.6603 & 7.1763 & 18.4803 $\pm$ 0.036 & 300 & HCT \\
& 5.6858 & 7.2018 & 18.5763 $\pm$ 0.029 & 300 & HCT \\
& 5.7054 & 7.2214 & 18.5873 $\pm$ 0.024 & 300 & HCT \\
& 5.7231 & 7.2391 & 18.5243 $\pm$ 0.029 & 300 & HCT \\
& 5.7507 & 7.2667 & 18.5683 $\pm$ 0.027 & 300 & HCT \\
& 5.7790 & 7.295 & 18.5873 $\pm$ 0.027 & 300 & HCT \\
& 5.7949 & 7.3109 & 18.5803 $\pm$ 0.036 & 300 & HCT \\
& 5.8130 & 7.329 & 18.6003 $\pm$ 0.027 & 300 & HCT \\
& 5.8454 & 7.3614 & 18.5893 $\pm$ 0.035 & 300 & HCT \\
& 5.8704 & 7.3864 & 18.6283 $\pm$ 0.037 & 300 & HCT \\
& 5.9002 & 7.4162 & 18.6643 $\pm$ 0.049 & 300 & HCT \\
& 6.5919 & 8.1079 & 18.8233 $\pm$ 0.039 & 300 & HCT \\
& 6.5995 & 8.1155 & 18.8373 $\pm$ 0.029 & 300 & HCT \\
& 6.6234 & 8.1394 & 18.8053 $\pm$ 0.036 & 300 & HCT \\
& 6.6849 & 8.2009 & 18.8453 $\pm$ 0.025 & 600 & HCT \\
& 6.6932 & 8.2092 & 18.8503 $\pm$ 0.030 & 600 & HCT \\
\hline
\end{tabular}
\end{center}
\end{table}
%%%                  split in V band %%%%
\begin{table}
\begin{center}
 {\bf Table 2.}~Optical Transient (contd.)\\
\begin{tabular}{ccccll} \hline 
&Date (UT) &Time since Burst& Magnitude & Exposure time & Telescope  \\
         &&in days&   (mag)&(Seconds)&  \\   \hline 
& 6.7327 & 8.2487 & 18.8503 $\pm$ 0.032 & 600 & HCT \\
& 6.7413 & 8.2573 & 18.8443 $\pm$ 0.022 & 600 & HCT \\
& 6.7792 & 8.2952 & 18.8363 $\pm$ 0.029 & 600 & HCT \\
& 6.7878 & 8.3038 & 18.8433 $\pm$ 0.034 & 600 & HCT \\
& 6.8205 & 8.3365 & 18.8523 $\pm$ 0.029 & 600 & HCT \\
& 6.8286 & 8.3446 & 18.8473 $\pm$ 0.047 & 600 & HCT \\
& 6.8851 & 8.4011 & 18.8823 $\pm$ 0.043 & 600 & HCT \\
& 6.8935 & 8.4095 & 18.8773 $\pm$ 0.038 & 600 & HCT \\
& 6.9016 & 8.4176 & 18.8183 $\pm$ 0.031 & 600 & HCT \\
& 7.5936 & 9.1096 & 18.897 $\pm$ 0.052 & 300 & ST \\
& 7.5980 & 9.114 & 18.945 $\pm$ 0.067 & 300 & ST \\
& 8.6070 & 10.123 & 19.017 $\pm$ 0.046 & 400 & ST \\
& 8.6483 & 10.1643 & 19.049 $\pm$ 0.041 & 600 & ST \\
& 8.6778 & 10.1938 & 19.064 $\pm$ 0.032 & 600 & ST \\
& 9.5890 & 11.105 & 19.078 $\pm$ 0.057 & 400 & ST \\
& 9.5905 & 11.1065 & 19.1263 $\pm$ 0.033 & 900 & HCT \\
& 9.6321 & 11.1481 & 19.0963 $\pm$ 0.037 & 900 & HCT \\
& 9.6875 & 11.2035 & 19.1403 $\pm$ 0.037 & 300X2 & HCT \\
& 9.7242 & 11.2402 & 19.1743 $\pm$ 0.039 & 300X3 & HCT \\
& 9.8502 & 11.3662 & 19.1663 $\pm$ 0.050 & 900 & HCT \\
& 10.6068 & 12.1228 & 19.3113 $\pm$ 0.030 & 900 & HCT \\
& 10.6157 & 12.1317 & 19.440 $\pm$ 0.095 & 400 & ST \\
& 10.6437 & 12.1597 & 19.2923 $\pm$ 0.052 & 900 & HCT \\
& 10.6483 & 12.1643 & 19.294 $\pm$ 0.057 & 900 & ST \\
& 10.6825 & 12.1985 & 19.2553 $\pm$ 0.050 & 300X3 & HCT \\
& 10.7272 & 12.2432 & 19.3483 $\pm$ 0.039 & 300+450 & HCT \\
& 10.7390 & 12.255 & 19.3503 $\pm$ 0.041 & 300 & HCT \\
& 10.7574 & 12.2734 & 19.3283 $\pm$ 0.049 & 900 & HCT \\
& 10.8689 & 12.3849 & 19.3463 $\pm$ 0.060 & 900 & HCT \\
& 11.6743 & 13.1903 & 19.4273 $\pm$ 0.038 & 300X3 & HCT \\
& 11.7064 & 13.2224 & 19.527 $\pm$ 0.154 & 600 & ST \\
& 11.7082 & 13.2242 & 19.2813 $\pm$ 0.053 & 300X2 & HCT \\
& 22.6028 & 24.1188 & 20.1233 $\pm$ 0.043 & 600 & HCT \\
& 22.6279 & 24.1439 & 20.0573 $\pm$ 0.040 & 600X2 & HCT \\
& 22.8033 & 24.3193 & 20.0633 $\pm$ 0.025 & 600X3 & HCT \\
& 23.6311 & 25.1471 & 20.0983 $\pm$ 0.032 & 300X5 & HCT \\
2003 & May&&&&\\
& 1.6069 & 33.1229 & $\pm$20.692 $\pm$ 0.090 & 2*300 & ST \\
\hline
\end{tabular}
\end{center}
\end{table}
%%%                  split in V band %%%%
\begin{table}
\begin{center}
 {\bf Table 2.}~Optical Transient (contd.)\\
\begin{tabular}{ccccll} \hline 
&Date (UT) &Time since Burst& Magnitude & Exposure time & Telescope  \\
         &&in days&   (mag)&(Seconds)&  \\   \hline 
%%%%    I band starts here%%%%
\multicolumn{5}{c}{\bf $I-$ passband}  \\
2003 & March&&&&\\
& 29.6306 & 0.1466 & 13.037 $\pm$ 0.026 & 200 & ST \\
& 29.6499 & 0.1659 & 13.185 $\pm$ 0.029 & 50 & ST \\
& 29.6855 & 0.2015 & 13.328 $\pm$ 0.029 & 50 & ST \\
& 29.6948 & 0.2108 & 13.487 $\pm$ 0.031 & 50 & ST \\
& 29.7083 & 0.2243 & 13.437 $\pm$ 0.030 & 100 & ST \\
& 29.7241 & 0.2401 & 13.501 $\pm$ 0.032 & 100 & ST \\
& 29.7374 & 0.2534 & 13.559 $\pm$ 0.032 & 100 & ST \\
& 29.7504 & 0.2664 & 13.618 $\pm$ 0.032 & 100 & ST \\
& 29.7616 & 0.2776 & 13.685 $\pm$ 0.032 & 200 & ST \\
& 29.7959 & 0.3119 & 13.837 $\pm$ 0.049 & 100 & ST \\
& 30.7698 & 1.2858 & 15.972 $\pm$ 0.023 & 200 & ST \\
& 30.7872 & 1.3032 & 15.985 $\pm$ 0.021 & 200 & ST \\
& 30.8046 & 1.3206 & 15.978 $\pm$ 0.021 & 200 & ST \\
& 30.8272 & 1.3432 & 16.058 $\pm$ 0.021 & 200 & ST \\
& 31.6683 & 2.1843 & 16.240 $\pm$ 0.034 & 100 & ST \\
& 31.6933 & 2.2093 & 16.290 $\pm$ 0.033 & 200 & ST \\
& 31.7110 & 2.227 & 16.305 $\pm$ 0.034 & 200 & ST \\
& 31.7338 & 2.2498 & 16.342 $\pm$ 0.031 & 200 & ST \\
& 31.7535 & 2.2695 & 16.329 $\pm$ 0.031 & 200 & ST \\
& 31.7715 & 2.2875 & 16.326 $\pm$ 0.033 & 200 & ST \\
& 31.7887 & 2.3047 & 16.341 $\pm$ 0.030 & 200 & ST \\
& 31.8088 & 2.3248 & 16.351 $\pm$ 0.030 & 200 & ST \\
& 31.8326 & 2.3486 & 16.387 $\pm$ 0.029 & 300 & ST \\
& 31.8539 & 2.3699 & 16.411 $\pm$ 0.028 & 300 & ST \\
& 31.8750 & 2.391 & 16.404 $\pm$ 0.028 & 300 & ST \\
& 31.8960 & 2.412 & 16.460 $\pm$ 0.028 & 300 & ST \\
2003 & April&&&&\\
& 1.6086 & 3.1246 & 16.591 $\pm$ 0.032 & 200 & ST \\
& 1.6890 & 3.205 & 16.586 $\pm$ 0.033 & 200 & ST \\
& 1.8294 & 3.3454 & 16.639 $\pm$ 0.029 & 300 & ST \\
& 1.8338 & 3.3498 & 16.631 $\pm$ 0.029 & 300 & ST \\
& 2.6738 & 4.1898 & 17.211 $\pm$ 0.031 & 300 & ST \\
& 2.7077 & 4.2237 & 17.224 $\pm$ 0.032 & 300 & ST \\
& 2.7572 & 4.2732 & 17.254 $\pm$ 0.035 & 300 & ST \\
& 4.6886 & 6.2046 & 17.586 $\pm$ 0.046 & 400 & ST \\
& 4.7180 & 6.234 & 17.663 $\pm$ 0.042 & 400 & ST \\
& 7.6029 & 9.1189 & 18.595 $\pm$ 0.098 & 300 & ST \\
& 8.5965 & 10.1125 & 18.674 $\pm$ 0.077 & 400 & ST \\
& 8.6405 & 10.1565 & 18.764 $\pm$ 0.052 & 600 & ST \\
& 8.6712 & 10.1872 & 18.614 $\pm$ 0.064 & 400 & ST \\
\hline
\end{tabular} 
\end{center} 
\end{table} 
%
%-------------------Table 3: ESO/SEST data------------
%
\begin{table}
\begin{center}
 {\bf Table 3.}~Results of the ESO/SEST scans on GRB 030329\\
\begin{tabular}{lccc}\hline
UT date of 2003 & freq (GHz) & flux (mJy)   & Beam \\
\hline
\hline
Mar 29/30       &   86       &  82 $\pm$ 52 & 57$^{\prime\prime}$ \\
                &  215       & 201 $\pm$ 308& 23$^{\prime\prime}$ \\
Mar 30/31       &   86       &  -2 $\pm$ 78 & 57$^{\prime\prime}$ \\
                &  215       & -49 $\pm$ 262& 23$^{\prime\prime}$ \\
Mar 31/Apr 1    &   86       &  38 $\pm$ 65 & 57$^{\prime\prime}$ \\
                &  215       &-144 $\pm$ 250& 23$^{\prime\prime}$ \\
\hline
\end{tabular} 
\end{center} 
\end{table} 
%
%------------------------Table 4 : IRAM/PdB Data----------------
%
\begin{table}
\begin{center}
 {\bf Table 4.}~Results of the  IRAM/PdB scans on GRB 030329 \\
\begin{tabular}{lcccc}\hline
UT date of 2003 & Config. & freq &  flux &  Beam and P.A. \\
&&(GHz)&(mJy)& \\ \hline
31 Mar 00:04 to 03:07 &  6Dp &  86.253 &  58.6 $\pm$ 0.5 &  14.1$^{\prime\prime}$ x 4.0$^{\prime\prime}$ at  55$^{\circ}$ \\
                      &              & 232.032 &  46.8 $\pm$ 3.1 &   5.2$^{\prime\prime}$ x 1.5$^{\prime\prime}$ at  55$^{\circ}$ \\
31 Mar 18:33 to 23:57 &  6Dp  &  98.473 &  58.2 $\pm$ 0.6 &   6.4$^{\prime\prime}$ x 4.2$^{\prime\prime}$ at  98$^{\circ}$ \\
                      &              & 238.500 &  40.6 $\pm$ 2.1 &   2.8$^{\prime\prime}$ x 1.6$^{\prime\prime}$ at  84$^{\circ}$ \\
01 Apr 20:20 to 21:55 &  6Dp  &  86.673 &  51.7 $\pm$ 0.9 &   9.1$^{\prime\prime}$ x 4.2$^{\prime\prime}$ at  98$^{\circ}$ \\  
                      &              & 240.528 &  23.8 $\pm$ 3.2 &   3.9$^{\prime\prime}$ x 1.6$^{\prime\prime}$ at 100$^{\circ}$ \\
05 Apr 16:54 to 18:04 &  6Dp  & 115.447 &  40.4 $\pm$ 3.7 &  10.3$^{\prime\prime}$ x 3.1$^{\prime\prime}$ at -53$^{\circ}$ \\
10 Apr 17:42 to 19:12 &  6Dp  &  86.244 &  23.5 $\pm$ 0.4 &  11.0$^{\prime\prime}$ x 4.2$^{\prime\prime}$ at -62$^{\circ}$ \\
                      &              & 232.171 &  12.8 $\pm$ 1.3 &   4.5$^{\prime\prime}$ x 1.5$^{\prime\prime}$ at -65$^{\circ}$ \\  
14 Apr 20:10 to 22:16 &  6Dp  &  91.333 &  14.9 $\pm$ 0.4 &   8.1$^{\prime\prime}$ x 4.2$^{\prime\prime}$ at  84$^{\circ}$ \\  
                      &              & 217.029 &   9.2 $\pm$ 2.2 &   3.6$^{\prime\prime}$ x 1.7$^{\prime\prime}$ at  80$^{\circ}$ \\ 
18 Apr 19:39 to 21:10 &  6Dp  & 115.271 &   7.7 $\pm$ 1.0 &   6.3$^{\prime\prime}$ x 3.3$^{\prime\prime}$ at  90$^{\circ}$ \\ 
                      &              & 232.032 &   6.5 $\pm$ 1.8 &   3.3$^{\prime\prime}$ x 1.5$^{\prime\prime}$ at  87$^{\circ}$ \\
24 Apr 17:36 to 19:32 &  6Dp  &  96.250 &   4.7 $\pm$ 0.7 &   8.2$^{\prime\prime}$ x 4.0$^{\prime\prime}$ at -67$^{\circ}$ \\ 
                      &              & 241.480 &   0.0 $\pm$ 3.8 &   3.5$^{\prime\prime}$ x 1.4$^{\prime\prime}$ at 105$^{\circ}$ \\ 
03 May 15:59 to 19:49 &  6Dp  &  86.243 &   2.9 $\pm$ 0.3 &   9.5$^{\prime\prime}$ x 4.4$^{\prime\prime}$ at 108$^{\circ}$ \\
                      &              & 233.467 &   1.4 $\pm$ 1.3 &   3.6$^{\prime\prime}$ x 1.6$^{\prime\prime}$ at -76$^{\circ}$ \\ 
16 May 13:20 to 15:57 &  5Dp  &  86.243 &   1.1 $\pm$ 0.8 &  12.4$^{\prime\prime}$ x 5.3$^{\prime\prime}$ at -54$^{\circ}$ \\ 
                      &              & 231.490 &   0.0 $\pm$ 10.1&   4.6$^{\prime\prime}$ x 1.9$^{\prime\prime}$ at -59$^{\circ}$ \\  
28 May 21:29 to 23:22 &  5Dp   &  84.443 &   0.4 $\pm$ 0.7 &  16.6$^{\prime\prime}$ x 5.5$^{\prime\prime}$ at  52$^{\circ}$ \\ 
                      &              & 238.500 &   0.0 $\pm$ 8.0 &   5.5$^{\prime\prime}$ x 1.9$^{\prime\prime}$ at  52$^{\circ}$ \\ 
20 June 12:41 to 17:49 & 4Dp &  95.434 &   0.6 $\pm$ 1.3 &  5.92$^{\prime\prime}$ x 5.63$^{\prime\prime}$ at  64$^{\circ}$ \\ 
\end{tabular} 
\end{center} 
\end{table} 
\clearpage
%
%------------------------FIGURES------------------------
%
\begin{figure}[h]\centering
\includegraphics[width=18.0cm]{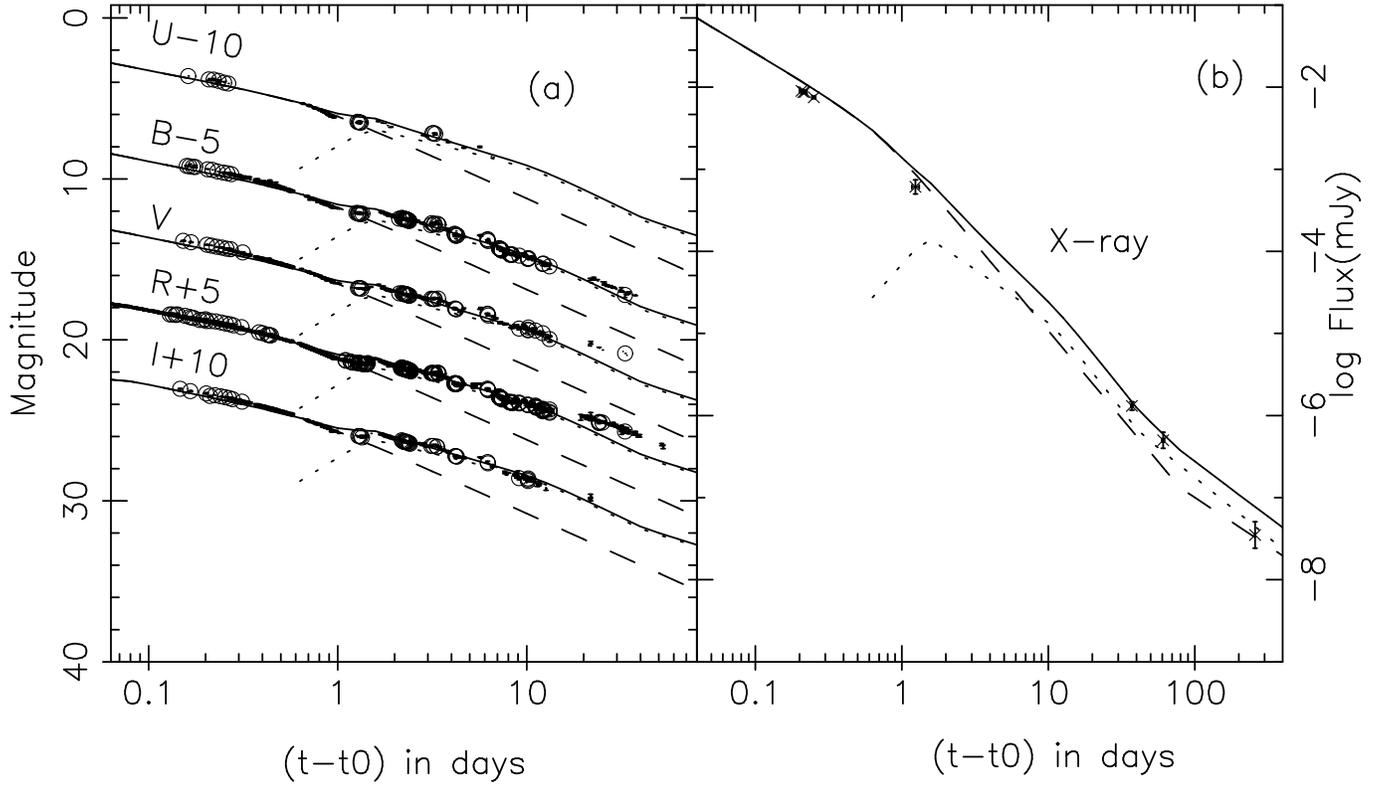}
\caption{(a) The Optical lightcurve of the afterglow of GRB030329. Open circles
represent the data presented in this paper and filled circles are those
from the literature (Lipkin et al. 2004). The solid line shows the total
flux predicted by model 1 discussed in the text, which has two jets as
in Berger et al. (2003). The dashed line shows the contribution of the
narrow jet alone and the dotted line that of the wide jet in each band.
(b) X-ray observations reported
by Tiengo et al. 2003 and 2004, shown along with the prediction of model 1.
Contribution of the narrow jet and the wide jet are shown separately as
the dashed and the dotted line respectively. The total flux is shown as the solid curve.}
\end{figure}
\begin{figure}[h]\centering
\includegraphics[width=18.0cm]{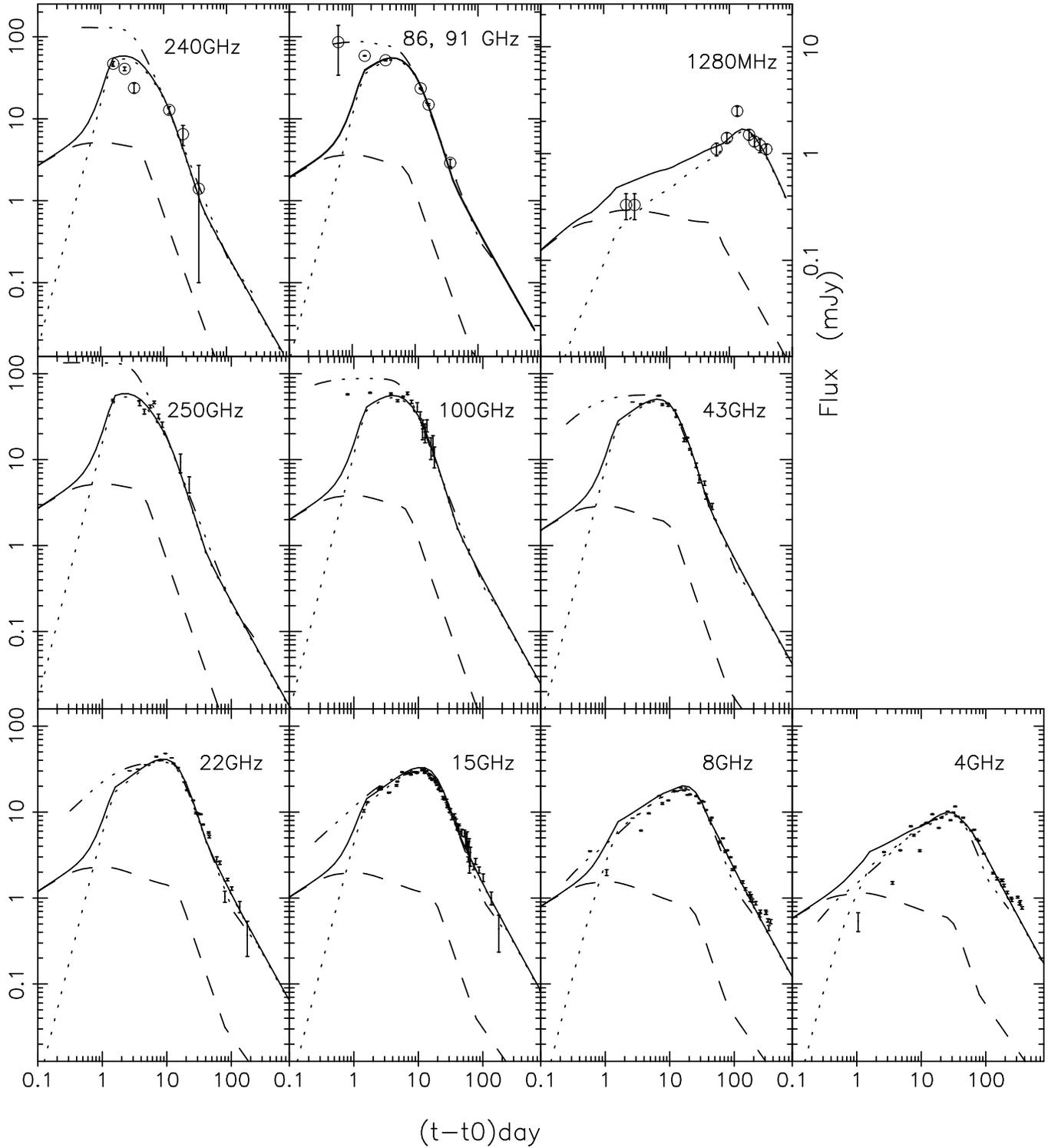}
\caption{Millimeter and radio observations of GRB030329 afterglow along with
the predictions of model 1 (two jets). Observations reported in this paper are represented
by open circles, crosses are data points from Berger et al. (2003) and
Sheth et al. (2003). The dashed and the dotted lines represent contributions of the narrow
and the wide jet respectively, the solid line shows the sum. The dash-dot-dot line
shows a model fit for an assumed stellar-wind density profile for the ambient medium.}
\end{figure}
\begin{figure}[h]\centering
\includegraphics[width=18.0cm]{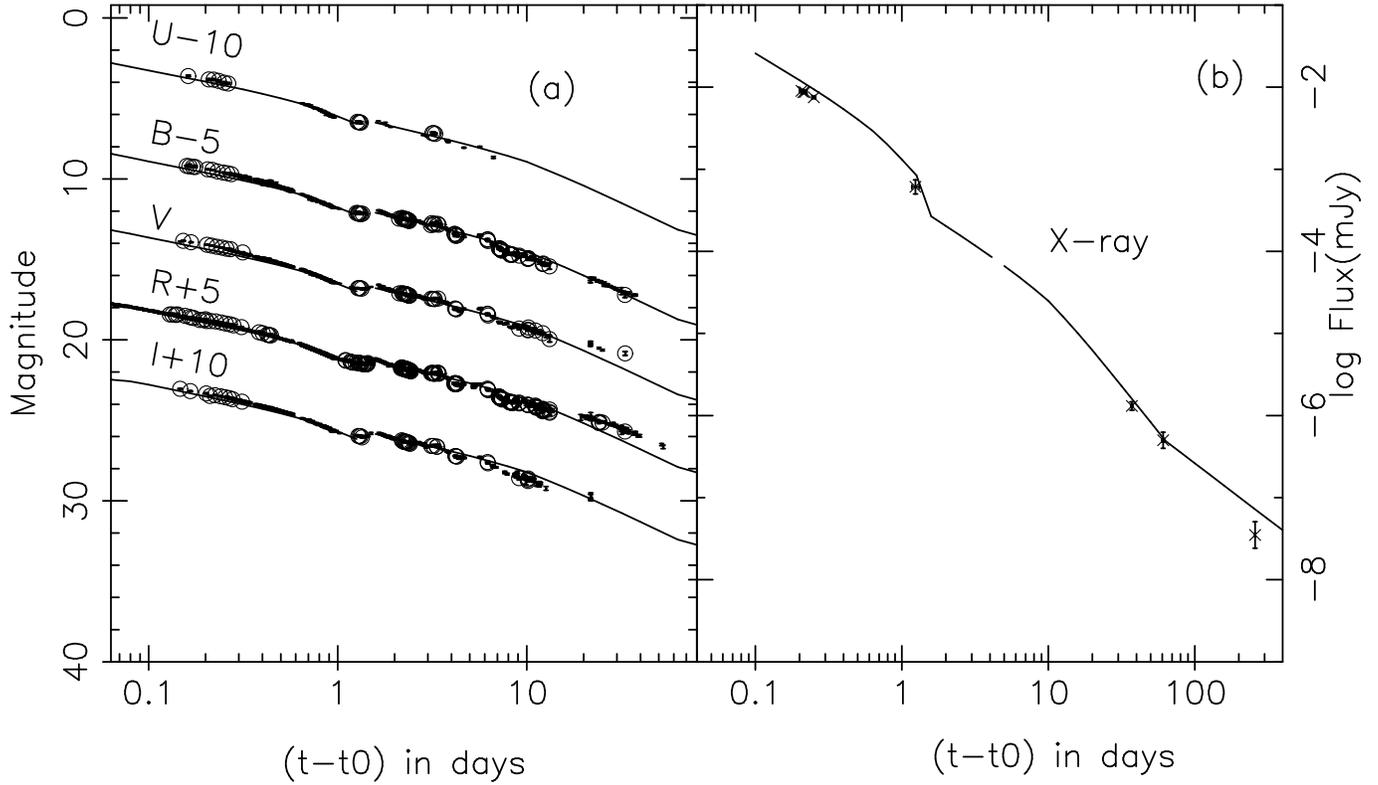}
\caption{(a) The Optical lightcurve of the afterglow of GRB030329, shown
with the prediction of model 2 (solid line), which assumes a transition of
an initially narrow jet to a wider jet
at $\sim 1.5$ days. (b) X-ray observations reported by Tiengo et
al.\ 2003 and 2004, with predictions of model 2. The flattening seen at late
times is due to the transition into non-relativistic regime at $\sim 63$ days.}
\end{figure}
\begin{figure}[h]\centering
\includegraphics[width=18.0cm]{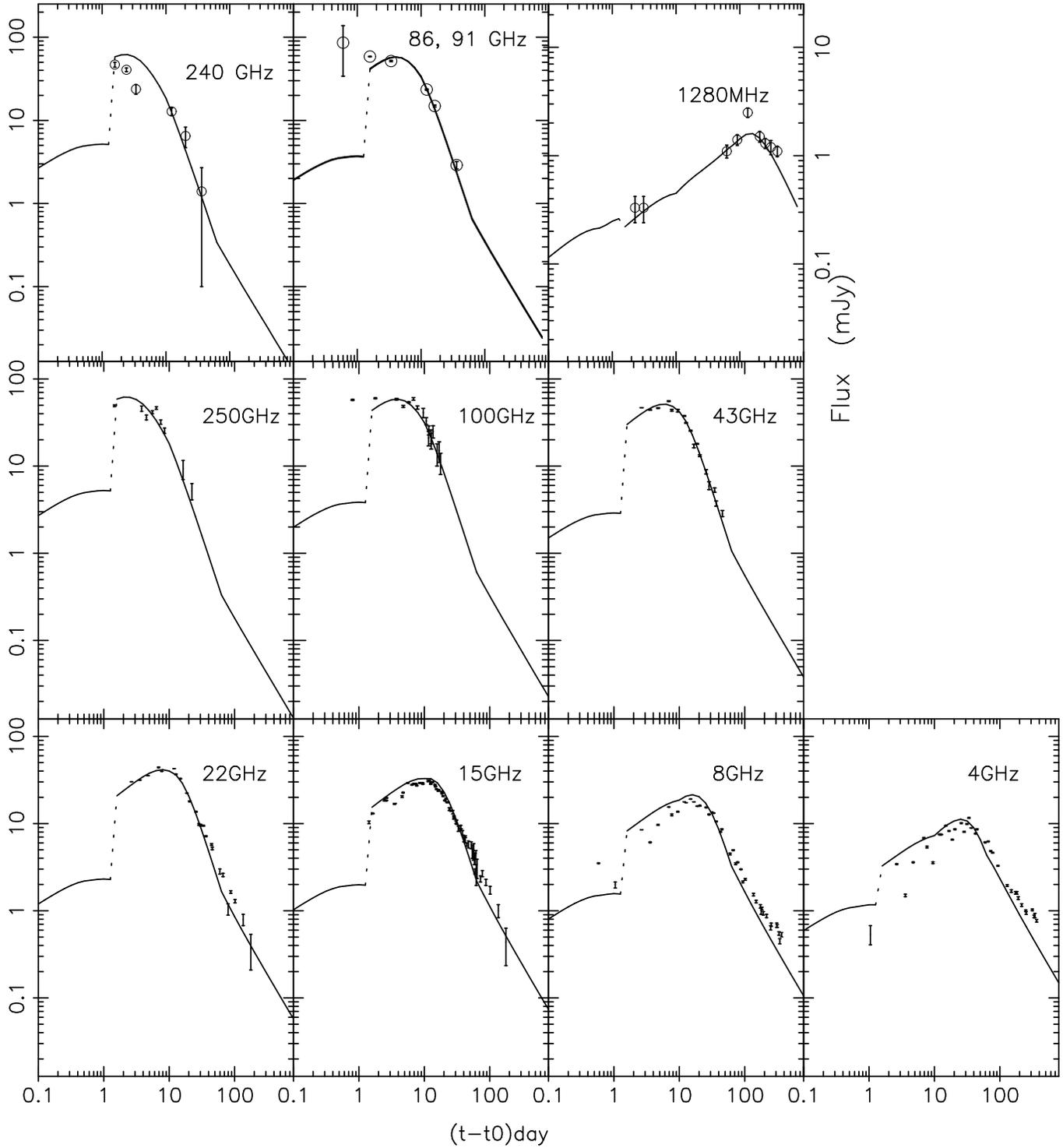}
\caption{Millimeter and radio observations of GRB030329, along with the predictions of
model 2 (refreshed jet) shown as the solid line. Open circles are data presented in this
paper, crosses represent data taken from Berger et al (2003) and Sheth et al (2003).}
\end{figure}
\begin{figure}[h]\centering
\includegraphics[width=14.0cm]{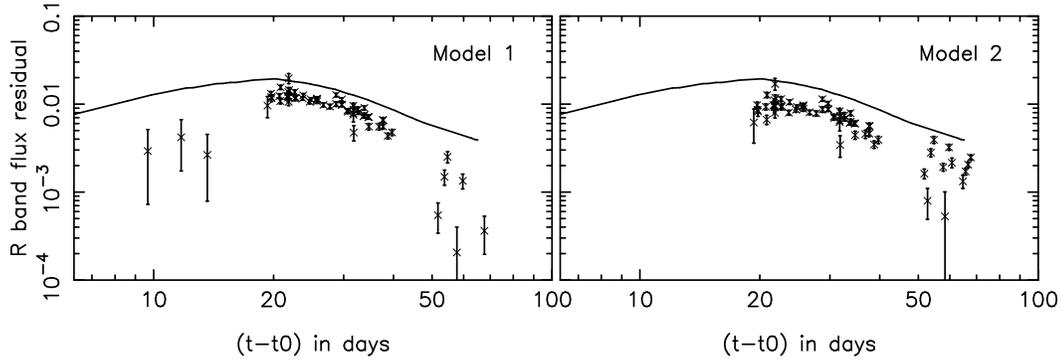}
\caption{$R$-band residuals for epochs beyond $\sim 7$ days, after subtracting
the modeled flux of the afterglow and the contribution of the host galaxy
($R=22.6$, Gorosabel et al. 2005b) from the observed flux. The two models
are shown in adjacent panels. This shows the R-band contribution needed from
the associated supernova SN2003dh to explain the total observed light from
the OT. The solid line is the red-shifted K-corrected SN1998bw R-band
lightcurve, shown for comparison.}
\end{figure}
\end{document}